\RequirePackage{fix-cm}
\documentclass[twocolumn,epjc3]{svjour3}  
\smartqed  
\RequirePackage{graphicx}
\usepackage[active]{srcltx}
\usepackage{amsmath}
\usepackage{amssymb}
\usepackage{array}

\newcommand{\be}{\begin{equation}}
\newcommand{\ee}[1]{\label{#1} \end{equation}}
\newcommand{\ba}{\begin{eqnarray}}
\newcommand{\ea}[1]{\label{#1} \end{eqnarray}}
\newcommand{\nl}{\nonumber \\}

\journalname{Eur. Phys. J. C}
\begin{document}

\title{Multiplicity Dependence of Hadron Spectra in Proton-Proton Collisions at LHC Energies and Super-Statistics
}


\author{Karoly Urmossy\thanksref{e1,addr1}
}

\thankstext{e1}{e-mail: karoly.uermoessy@cern.ch}


\institute{Institute for Particle and Nuclear Physics, Wigner RCP of the HAS\newline 29-33 Konkoly--Thege Mikl\'os Str., H-1121 Budapest, Hungary\label{addr1}
}

\date{Received: date / Accepted: date}

\maketitle

\begin{abstract}
In this paper, transverse momentum spectra of $\pi^+$, $K^+$ and $p$ measured at fix event-multiplicities and $\sqrt s$ = 0.9, 2.76 and 7 TeV collision energies by the CMS Collabotaion are shown to fit the Tsallis-distribution. It is found that the power of the distribution shows a double-logarithmic dependence on the event-multiplicity $N$, while the $T$ parameter depends linearly on $N$. A similar double-logarithmic dependence of the $q$ parameter of $\pi^0$ spectra on the collision energy $\sqrt{s}$ is found too.

It is also shown that event-by-event fluctuations of the multiplicity $N$ and the total $E_T$ energy going into the transverse region can be the reason for the emergence of the Tsallis distribution in high-energy proton-proton collisions.
\end{abstract}

\section{Introduction}
\label{intro}
Cut-power law hadron distributions (sometimes refered to as Tsallis distribution) are observed in various high-energy collisions: in electron-positron $e^+e^-$ \cite{bib:Beck1}~--~\cite{bib:Biro_eeFF}, proton-(anti-)pro\-ton $pp$ \cite{bib:Wilk1}~--~\cite{bib:UKppFF}, elastic $pp$ \cite{bib:Fagundes} and nucleus-nucleus $AA$ \cite{bib:Wilk2}~--~\cite{bib:Wilk7} collisions. While the dependence of the parameters of this distribution on the collision energy $\sqrt{s}$ \cite{bib:Beck1,bib:Beck2,bib:UKeeFF,bib:Wilk1,bib:HotQuarks2010,bib:Wibig,bib:UKppFF,bib:Wilk2,bib:Wilk3,bib:Shao,bib:Chincellato,bib:Cley4}, centrality \cite{bib:Shao,bib:Chincellato,bib:De,bib:Wilk7}, number of colliding nucleons $N_{coll}$ \cite{bib:De}, hadron species \cite{bib:UKeeFFstrange,bib:Biro_eeFF,bib:Cley1,bib:Cley2,bib:Cley3,bib:Tang1,bib:Shao,bib:UKinAuAu,bib:Tang2,bib:BiroEPJA40} and momentum transfer \cite{bib:Fagundes} is thoroughly discussed in the literature, the dependence of hadron spectra on the event-multiplcity measured in \cite{bib:CMS} has not been ana\-lysed yet. In Sec.~\ref{sec:spec}, this task is fulfilled.

One explanation for the emergence of cut-power laws may be found in thermal hadronisation models. These models are based on the conjecture that in high-energy collisions, small thermal droplets of matter (often refered to as ``fireballs'' or ``clusters'') are created and these droplets fragment into hadrons. The calculations are carried through either in the canonical \cite{bib:Wibig2,bib:Wibig,bib:OldBoys,bib:Becattini1,bib:Becattini3,bib:Becattini4} or microcanonical \cite{bib:Becattini10,bib:Becattini11,bib:Liu1,bib:Begun,bib:Becattini12,bib:Becattini13,bib:Liu2} framework, and describe measured data on hadron specra, total hadron multiplicities as well as multiplicity distributions. The latter can be approximated by either the negative binomial or by Euler's gamma-distribution.

Since each formed fireball is different, they may carry different four-momenta, fragment into different number of hadrons and - if one works in the canonical framework - they may have different temperature. It has been shown that, if a single cluster or fireball is described by the Boltzmann-Gibbs distribution, however, the temperature \cite{bib:Beck1,bib:Beck2,bib:Wilk2,bib:Wilk3,bib:Wilk4,bib:Wilk5,bib:Wilk6,bib:Wilk7} or cluster volume \cite{bib:Wilk5,bib:Begun} or the multiplicity of hadrons stemming from a cluster \cite{bib:UKeeFF,bib:UKppFF} fluctuates from cluster to cluster, the average hadron spectrum may become a cut-power law (or Tsallis) distribution. The case of volume \cite{bib:Begun} and multiplicity \cite{bib:UKeeFF,bib:UKeeFFstrange,bib:UKppFF} fluctuations has also been generalised to the microcanonical ensemble. The latter has been applied to the fragmentation of jets produced in $e^+e^-$ \cite{bib:UKeeFF,bib:UKeeFFstrange} and $pp$ collisions \cite{bib:UKppFF}.

It is worth noting that if we approximate very narrow jets with one-dimensional bunches of massless had\-rons, the conservation of four-momentum reduces to the conservation of energy, and the clusters (jets) become massless too (for a discussion, see \cite{bib:UKppFF}). Thus, jet-fragmentation models \cite{bib:UKeeFF,bib:UKeeFFstrange,bib:UKppFF} may be considered as an approximation of microcanonical statistical hadronisation models \cite{bib:Becattini10,bib:Becattini11,bib:Liu1,bib:Begun,bib:Becattini12,bib:Becattini13,bib:Liu2} for very high hadron energies and jet-like, very narrow, one-dimensional clusters. The latter models are sensitive to the species of hadrons and multiplicity distributions are obtained from them as a consequence. In the previous models, the multiplicity distribution is put in by hand but, because of their simplicity, analitic calculations are possible.

In this paper, I will focus on the effect of cluster energy fluctuations. Hadronic transverse momentum spectra in $pp$ collisions are mixtures of particle yields comming from many clusters of different energy $E_c$, four-velocity $u_\mu$, and yielding different $N$ number of hadrons. For the spectra published in \cite{bib:CMS} is soft, I will work in the canonical ensemble, supposing that each cluster produces hadrons according to an isotrope Boltzmann-Gibbs distribution in $D=3$ dimensions in the frame co-moving with the cluster:

\be
\left. \frac{dN}{d^3 \vec{p}} \right|_{\textrm{1\,cluster}} \sim\,\exp\left\lbrace -\beta (u_\mu p^\mu-m) \right\rbrace \;,
\ee{int0}
with $\beta$ being the inverse temperature and $p^\mu$ being the four-momentum of the hadron.

In Sec.~\ref{sec:Efluct}, I show that an appropriate choice for fluctuations of the cluster energy result in cut-power law (or Tsallis) shaped hadron spectrum. As a first approximation, I will neglect the cluster velocities choosing $u_\mu = (1,0,0,0)$ and do calculations for massless ($m=0$) particles. This way, $\beta = DN/E_c$ holds. Furtheremore, I will assume that only one cluster is formed in one $pp$ event.

In Sec.~\ref{sec:spec}, I show that the Tsallis distribution obtained in Sec.~\ref{sec:Efluct} fits spectra of identified $\pi^+$, $K^+$ and $p$ at fixed event multiplicity at LHC energies.

In Sec.~\ref{sec:ENfluct}, I show that multiplicity fluctuations of the Euler-gamma type superimposed over cluster energy fluctuations also result in an approximately Tsallis--shaped transverse $\pi^+$ spectrum and fit measurements well. In addition, I also show that a double-logarithmic dependence of the $q$ parameter of $\pi^0$ spectra on the collision energy $\sqrt s$ holds in the $\sqrt s \in$ [0.2, 7] TeV range.

In Sec.~\ref{sec:pEN}, I make a prediction on the joint distribution of the event multiplicity $N$ and total transverse energy $E_T$ in an event, supposing that $E_T \propto E_c$.

\section{Power-law Spectrum from Cluster Energy Fluctuations}
\label{sec:Efluct}

According to Eq.~(\ref{int0}), let us conjecture that hadrons stemming from a cluster have an isotropic Boltzmann-Gibbs distribution in the frame co-moving with the cluster:

\be
\left. \frac{dN}{d^D \vec{p}} \right|_{\textrm{1\,cluster}} = f_{N,E_c}(\epsilon) = A\, \exp\left\lbrace -\frac{DN}{E_c}\epsilon \right\rbrace \;,
\ee{sup1}
with $E_c$ and $N$ being the energy and multiplicity of the cluster and $A = \left[DN/E_c \right]^D/[\kappa_D \Gamma(D)]$ comes from the normalisation condition

\be
\int d^D p\, f_{N,E_c}(\epsilon) = 1,
\ee{sup2}
where $\kappa_D = \int d\Omega_D$ is the angular part of the $D$ dimensional momentumspace integral. The masses of the hadrons have been neglected, taking $\epsilon=|\vec{p}|$. Let us also conjecture cluster energy fluctuations of the form

\be
g_N(E_c) = \frac{1}{\Gamma(\alpha+1)} \frac{\alpha E_0}{E^2_c} \left(\frac{\alpha E_0}{E_c} \right)^{\alpha} e^{-\alpha E_0/E_c}\;.
\ee{sup4}

This way, the hadron distribution in a cluster of multiplicity $N$, averaged over fluctuations of the cluster energy, becomes a cut-power law (or Tsallis) distribution:

\ba
\left\langle\left. \frac{dN}{d^D \vec{p}} \right|_{\textrm{1\,cluster}} \right\rangle_{E_c} &=& f_N(\epsilon) =  \int dE_c \,g_N(E_c)\,f_{N,E_c}(\epsilon)\nl
&=& A_N \left( 1+\frac{q-1}{T}\epsilon \right)^{-1/(q-1)}
\ea{sup6}
with

\ba
q &=& 1 + \frac{1}{\alpha+D+1} \;,\nl
T &=& \frac{\alpha E_0}{ DN (\alpha+D+1)} \;.
\ea{sup6.2}
The parameters $\alpha$ and $E_0$ (and thus $q$ and $T$) may depend on the hadron multiplicity $N$ in the cluster.

\section{Hadron Spectra at Fixed Multiplicity}
\label{sec:spec}
In this section, I show that Eq.~(\ref{sup6}) (in $D=3$ dimensions) fits transverse momentum spectra of various identified hadrons if the kinetic energy, $\epsilon \rightarrow m_T-m$, is used as scaling variable:

\be
\left. \frac{dN}{dp_T\, dy}\right|^{N=\textrm{fix}}_{y=0} = \frac{A\, p_T\, m_T} {\left[1 + \frac{q-1}{T} (m_T-m) \right]^{1/(q-1)}} .
\ee{spec1}
The analysed data are: transverse momentum spectra of $\pi^+$ (Fig.~\ref{fig:pipdNdpT}) $K^+$ (Fig.~\ref{fig:KpdNdpT}) and $p$ (Fig.~\ref{fig:ppdNdpT}) stemming from $pp$ collision events of fixed multiplicities (indicated as $N_{tracks}$ in the figures) at $\sqrt s$ = 0.9, 2.76 and 7 TeV collision energies and from the rapidity range $|y|\leq1$. Figs.~\ref{fig:pipq},~\ref{fig:Kpq},~\ref{fig:ppq} and \ref{fig:T} show that a

\ba
q &=& 1 + \mu\ln \ln (N/N_q) \;,\nl
T &=& T_0 (1 + N/N_T)
\ea{spec2}
dependence of the $q$ and $T$ parameters on the event-multiplicity $N$ is consistent with measurements. The parameters $\mu$, $N_q$, $T_0$ and $N_T$ take different values for each particle species, but do not seem to change significantly (within errors) as the collision energy $\sqrt s$ varies. It is interesting that while the $T$ parameters of $K^+$ and $p$ grow with $N$, the $T$ of $\pi^+$ is approximatly independent of the event-multiplicity. Since the dominant part of the produced hadrons is pions, we may use Eq.~(\ref{spec2}) to estimate the dependence of the $q$ and $T$ parameters of pions on the pion-multiplicity $N_\pi$, by substituting $N\rightarrow N_\pi$. However, it is perhaps not true for kaons and protons.

\section{The Effect of Multiplicity Fluctuations}
\label{sec:ENfluct}

\begin{figure}
\includegraphics[width=0.47\textwidth,height=0.32\textheight]{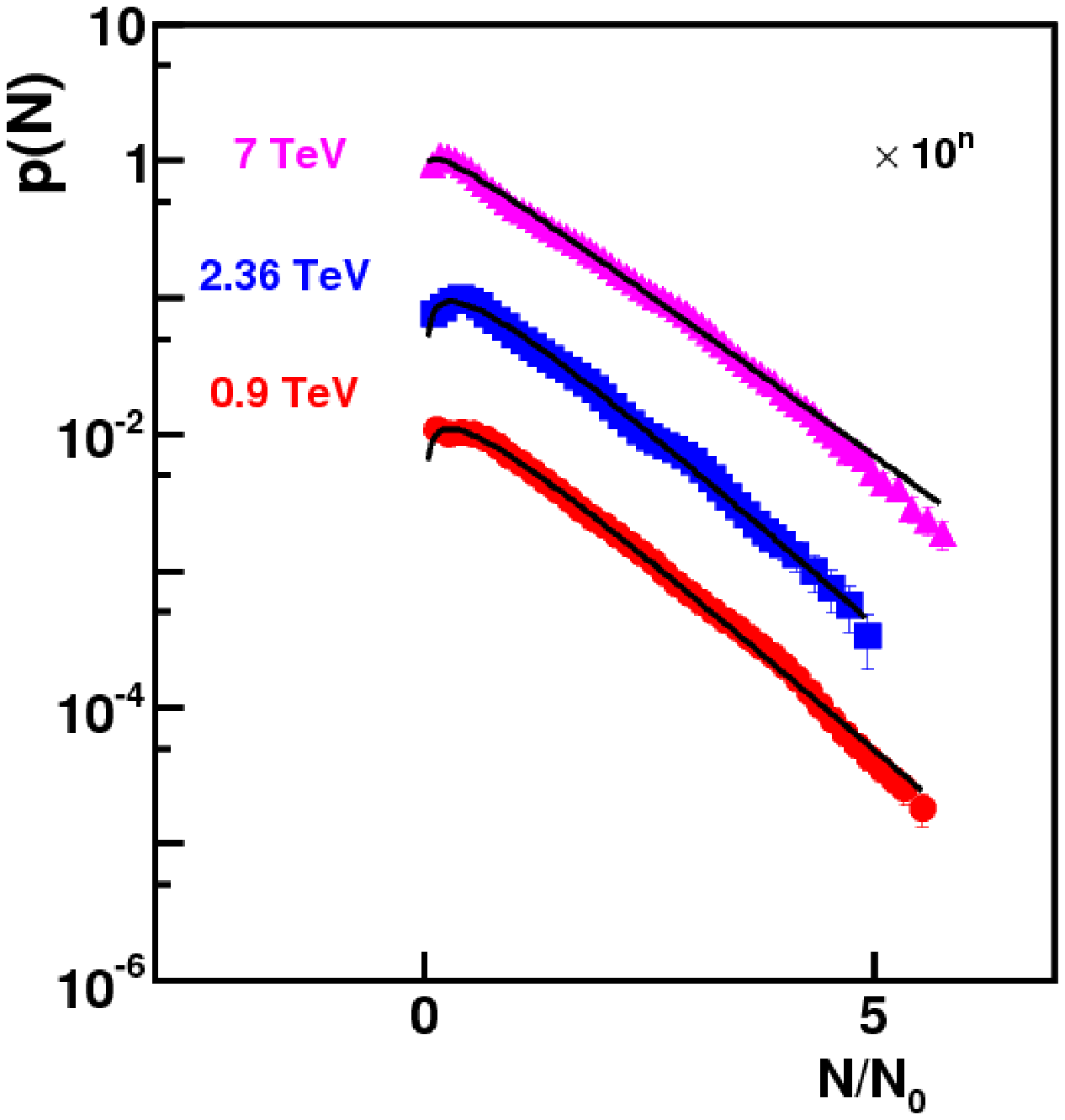}
\includegraphics[width=0.47\textwidth,height=0.32\textheight]{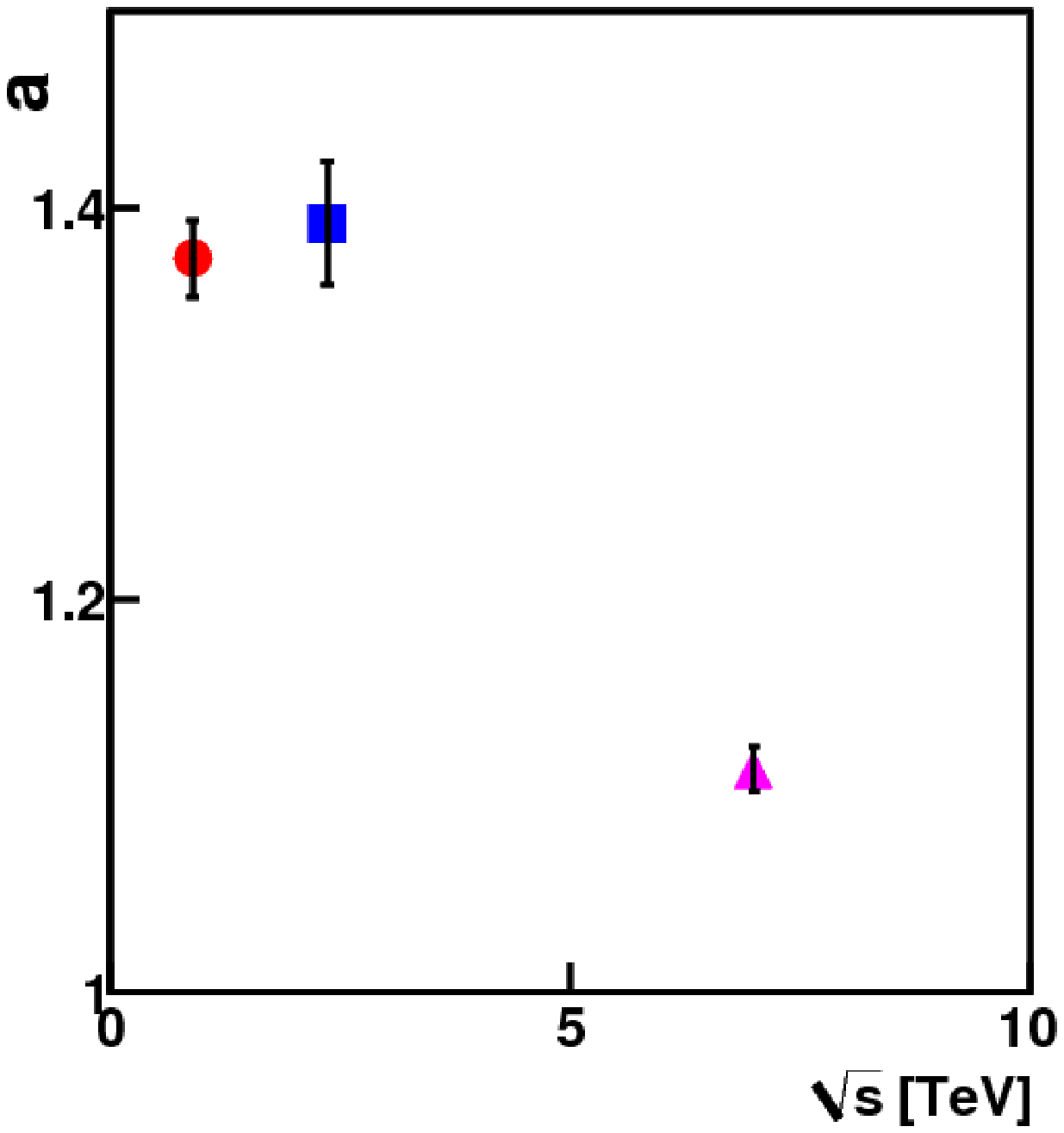}
\includegraphics[width=0.47\textwidth,height=0.32\textheight]{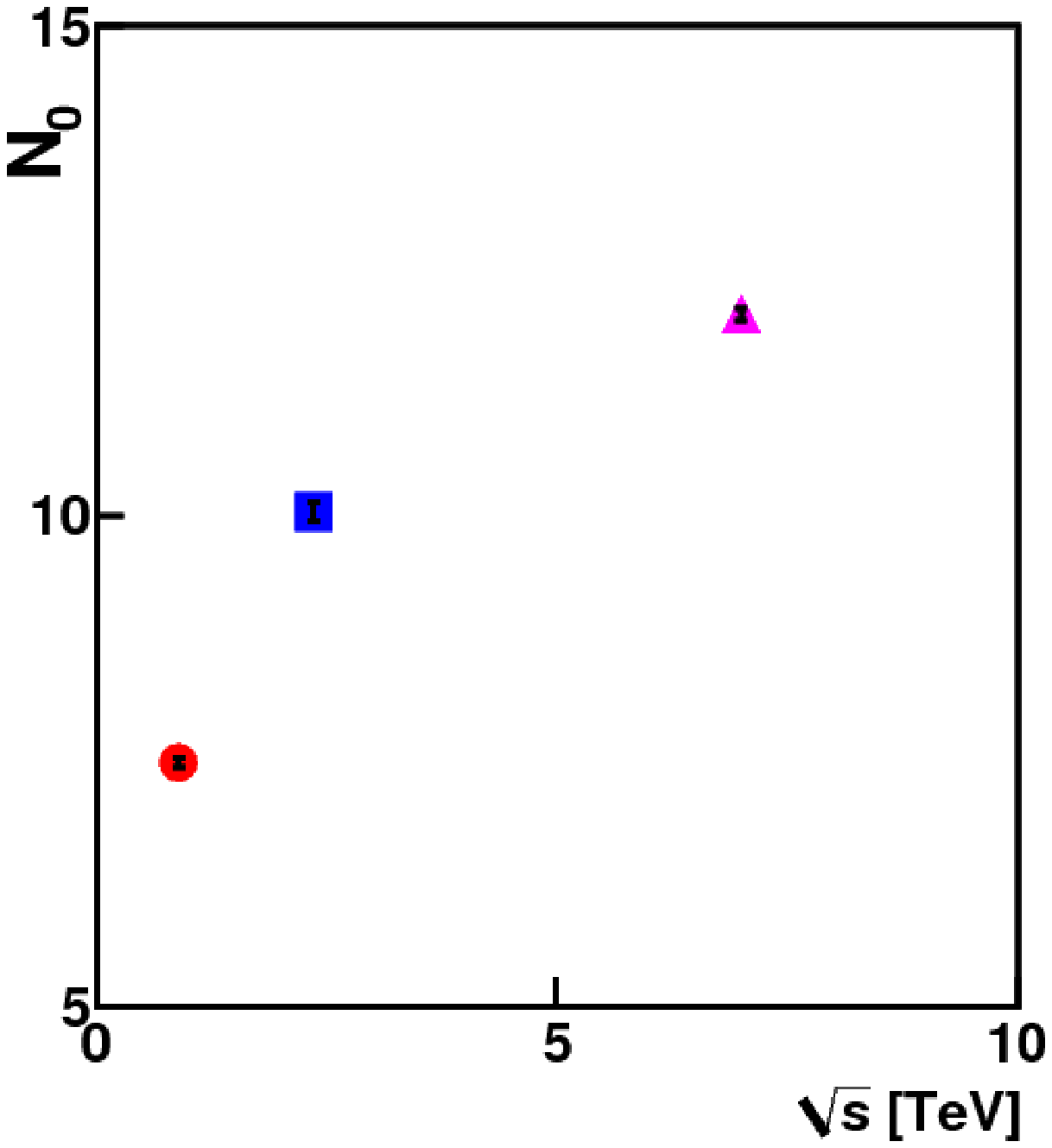}
\caption{Top: fits of Eq.~(\ref{sup5}) to multiplicity distributions of charged hadrons stemming from $pp$ collisions at $\sqrt s$ = 0.9, 2.76 and 7 TeV \cite{bib:ALICE1,bib:ALICE2}. Middle and bottom: dependence of the distribution's parameters on the collision energy. \label{fig:pN}}
\end{figure}

It has been known for a time that multiplicity distributions of charged hadrons in high-energy collisions may be approximated by the negative binomial (NBD) or Euler' Gamma distribution \cite{bib:Becattini10,bib:UKeeFF,bib:Becattini11,bib:UKppFF,bib:Begun,bib:Becattini12,bib:Becattini1,bib:Ghosh,bib:Dumitru}. Fig.~\ref{fig:pN} shows fits of Euler's Gamma-distribution

\be
p(N) = \frac{1}{\Gamma(a)}\frac{a}{N_0}\left(\frac{aN}{N_0} \right)^{a-1}e^{-aN/N_0}
\ee{sup5}
to multiplicity distributions measured by the ALICE Collaboration \cite{bib:ALICE1,bib:ALICE2} in the $|\eta|\leq1$ rapidity interval. (Eq.~(\ref{sup5}) is the $n\gg1$ and $k/\langle n \rangle \gg1$ limit of the NBD used in \cite{bib:Dumitru,bib:Ghosh} to analyse data in \cite{bib:ALICE1,bib:ALICE2}.)

The hadron distribution in a cluster, averaged over the fluctuations of both the cluster energy $E_c$ and the cluster multiplicity $N$, is

\be
\left\langle\left. \frac{dN}{d^D \vec{p}} \right|_{\textrm{1\,cluster}} \right\rangle_{E_c,N} = f(\epsilon) =  \sum_N  \,p(N)\,f_{N}(\epsilon) \;.
\ee{sup8}

For pions, the $m=0$ choice used in Sec.~\ref{sec:Efluct} may be an acceptible approximation in the measured $p_T$ range in \cite{bib:CMS}. Furtheremore, the multiplicity distribution of pions may take a similar form as the experimentally accessible multiplicity distributions of charged particles, since most of the hadrons produced in $pp$ collisions are pions. The analysis of transverse momentum spectra of pions in Sec.~\ref{sec:spec} show that the $q$ and $T$ parameters of the spectrum Eq.~(\ref{spec1}) depend weakly on the event multiplicity ($T\approx$ 70~-~80 MeV almost independently of $N$ and  $\mu\approx$ 0.13~-~0.14 in Eq.~(\ref{spec2})). Thus, the multiplicity and cluster energy averaged spectrum Eq.~(\ref{sup8}), evaluated with parameters obtained from the analysis of pion spectra in Sec.~\ref{sec:spec}, is close to the Tsallis distribution.

Fig.~\ref{fig:Naverpip} shows measured and calculated multiplicity-averaged $\pi^+$ spectra. In the evaluation of Eq.~(\ref{sup8}), the discrete sum was approximated by an integral with a lower bound $N_- = e\,N_q$ in order to ensure that $q\geq1$. For the $a$ parameter of the multiplicity distribution Eq.~(\ref{sup5}), I used values obtained from fits to experimental data shown in Fig.~\ref{fig:pN}. For the mean multiplicity parameter $N_0$, values higher then the ones obtained from fits to measured multiplicity distributions turned out to give best agreement with measured spectra (see the caption of Fig.~\ref{fig:Naverpip}).

It is also worth to note that the $q$ parameter of the transverse momentum spectra of neutral pions (which is also an $E_c$ and $N$ averaged observable) also show double-logarithmic dependence on the collision energy for $\sqrt s\in$ [0.2, 7] TeV

\be
q(s) = 1 + q_1\ln \ln (\sqrt{s}/Q_0) \;.
\ee{spec_1}
In the meanwhile, the $T$ parameter is not affected signi\-fi\-cant\-ly (within errors) by the grows of $s$. Fig.~\ref{fig:pi0} shows fit results. Similar functional forms for $q(s)$ have also been proposed in \cite{bib:Wibig2,bib:Wibig,bib:Wilk3,bib:Wilk4}.

From the observation that in $pp$ collisions at $\sqrt s$ = 900 GeV the $q$ parameter of various hadron species coincide within errors \cite{bib:Cley1,bib:Cley2,bib:Cley3}, we may infere that Eq.~(\ref{spec_1}) might hold for other hadons too.

\section{Joint Distribution of the Multiplicity and the Total Transverse Energy in an Event}
\label{sec:pEN}

In the model presented above, the joint distribution of cluster energy $E_c$ and cluster multiplicity $N$ is 

\be
p(N,E_c) = p(N) \, g_N(E_c) \;,
\ee{pNE}
where $p(N)$ is the $E_c$~-~averaged multiplicity distribution

\be
p(N) = \int dE_c\, p(N,E_c)
\ee{sup3.1}
shown in Fig.~\ref{fig:pN}, and $g_N(E_c)$ is the normalised distribution of $E_c$ at fix multiplicity. From Eqs.~(\ref{sup6.2}) and (\ref{spec2}), the parameters of $g_N(E_c)$ depend on $N$ as

\ba
\alpha &=& \frac{1}{\mu\ln\ln(N/N_q)} - D - 1 \;,\nl
E_0 &=& \frac{DNT_0\,(1+N/N_T)}{ 1 - (D+1)\mu\ln\ln(N/N_q)} \;.
\ea{sup6.3}

Since the hadron distribution in a cluster is assumed to be isotropic, the $E_T$ energy going into the transverse region ($|y|\leq1$ in \cite{bib:CMS}) in a sigle $pp$ event is proportional to the energy of the cluster formed in the event: $E_T \sim E_c$. Thus, the distributions of $E_T$ and $E_c$ have the same form. Fig.~\ref{fig:pNE} shows the joint distribution $p(N,E_T)$ with its projections, using parameters obtained from data on $\pi^+$ spectra measured at $\sqrt s$ = 7 TeV and in the rapidity range $|y|\leq1$.

\section{Conclusions}
\label{sec:conc}
Cut-power law hadron distributions (sometimes refered to as Tsallis distribution) are observed in various high-energy collisions (from electron--positron to nucleus--nucleus reactions \cite{bib:Beck1}~--~\cite{bib:Wilk7}). One possible explanation for this phenomena can be found in thermal hadronisation models. According to these models, in high-energy collisions, small thermal ``fireballs'' or ``clusters'' are formed and fragment into hadrons. Hadrons stemming from a single cluster inherit the thermal (canonical \cite{bib:OldBoys,bib:Becattini1,bib:Becattini3,bib:Becattini4} or microcanonical \cite{bib:Becattini10,bib:Becattini11,bib:Liu1,bib:Begun,bib:Becattini12,bib:Becattini13,bib:Liu2}) distribution of the fireball. However, since the four-momentum (or four-velocity) and cluster mass \cite{bib:Becattini10,bib:Becattini11,bib:Liu1,bib:Begun,bib:Becattini12,bib:Becattini13,bib:Liu2}, or the temperature \cite{bib:Beck1,bib:Beck2,bib:Wilk2,bib:Wilk3,bib:Wilk4,bib:Wilk5,bib:Wilk6,bib:Wilk7} or cluster volume \cite{bib:Wilk5,bib:Begun} or the hadron multiplicity \cite{bib:UKeeFF,bib:UKppFF} fluctuates from cluster to cluster, the average hadron spectrum may become a cut-power law (or Tsallis) distribution.

\begin{figure}[p]
\includegraphics[width=0.47\textwidth,height=0.32\textheight]{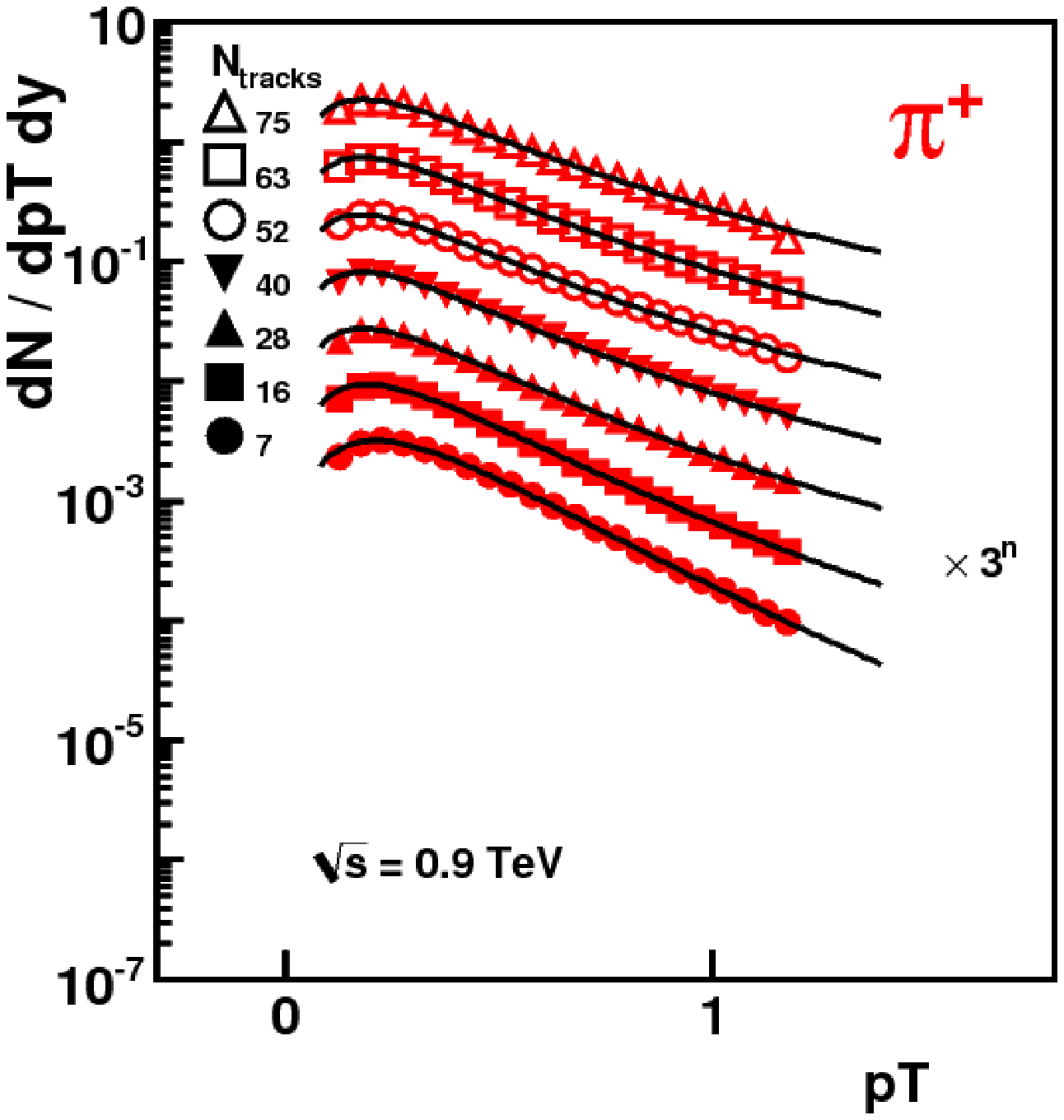}
\includegraphics[width=0.47\textwidth,height=0.32\textheight]{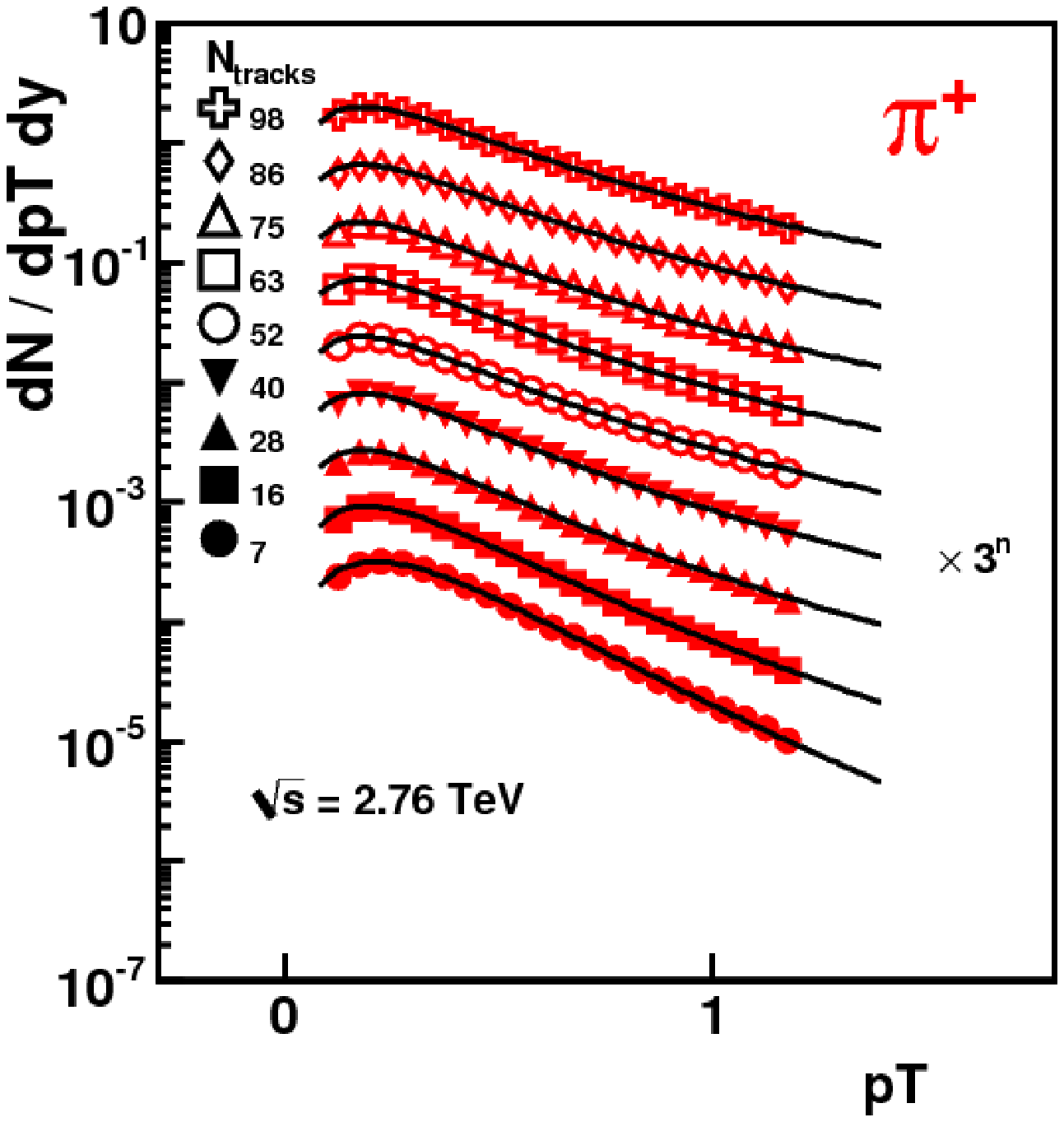}
\includegraphics[width=0.47\textwidth,height=0.32\textheight]{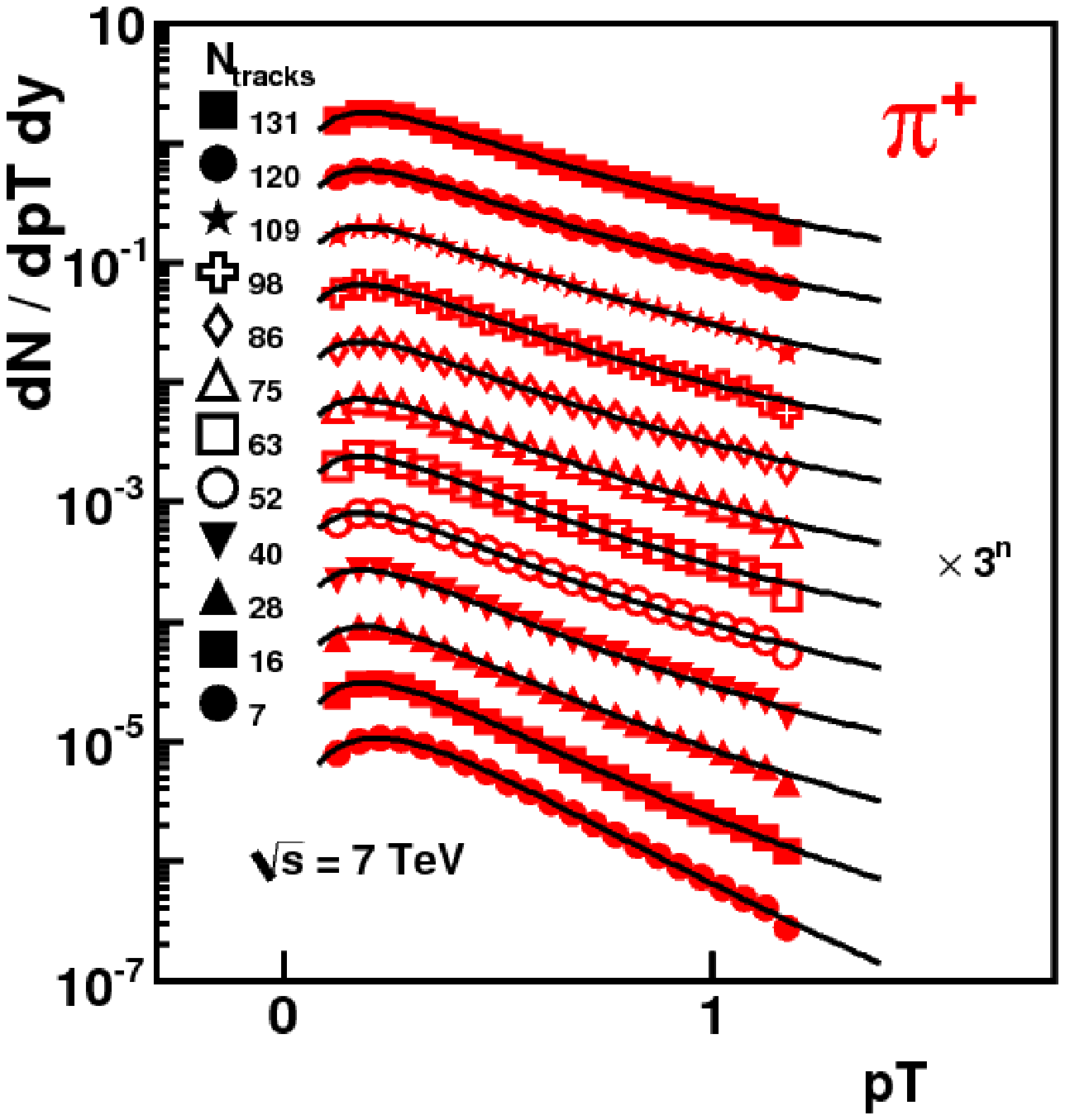}
\caption{Transverse momentum spectra of $\pi^+$-s stemming from $pp$ collisions of fix multiplicities, measured at $\sqrt{s}$ = 0.9 TeV (top), 2.76 TeV (middle) and 7 TeV (bottom) collision energies. Rapidity range: $|y|\leq1$. Data of graphs were published in \cite{bib:CMS}. Curves are fits of Eq.~(\ref{spec1}). \label{fig:pipdNdpT}}
\end{figure}

\begin{figure}
\includegraphics[width=0.47\textwidth,height=0.32\textheight]{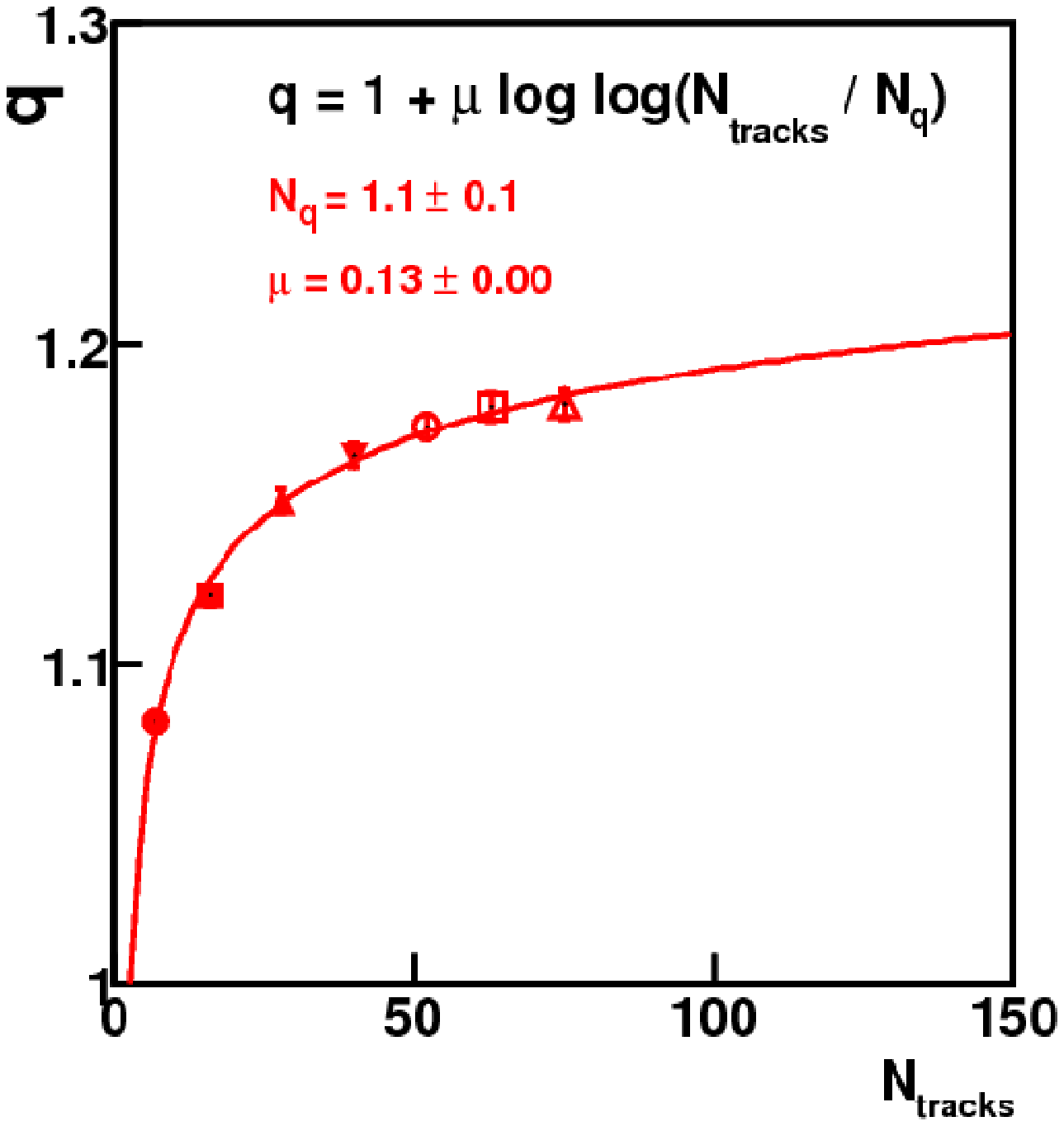}
\includegraphics[width=0.47\textwidth,height=0.32\textheight]{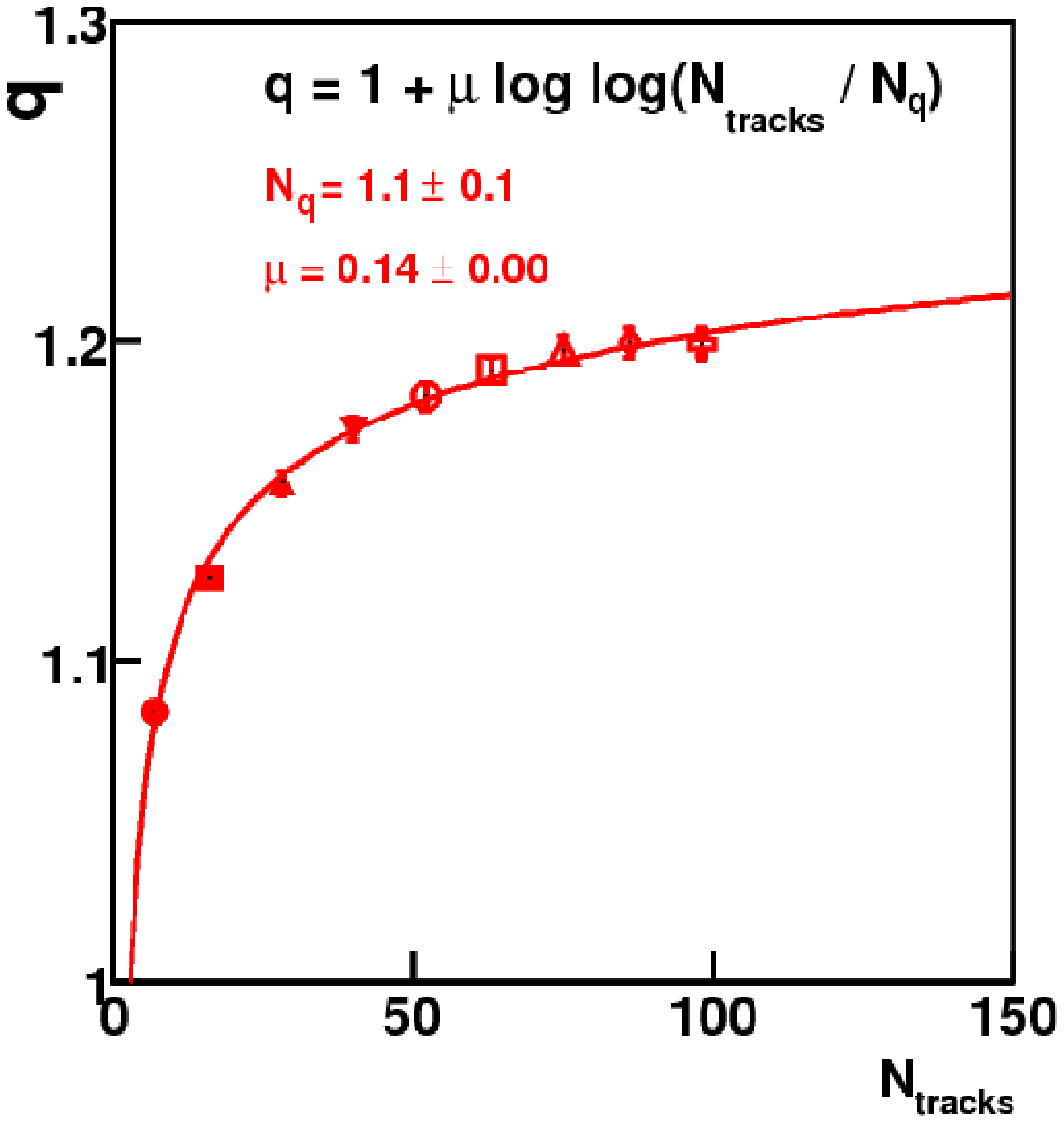}
\includegraphics[width=0.47\textwidth,height=0.32\textheight]{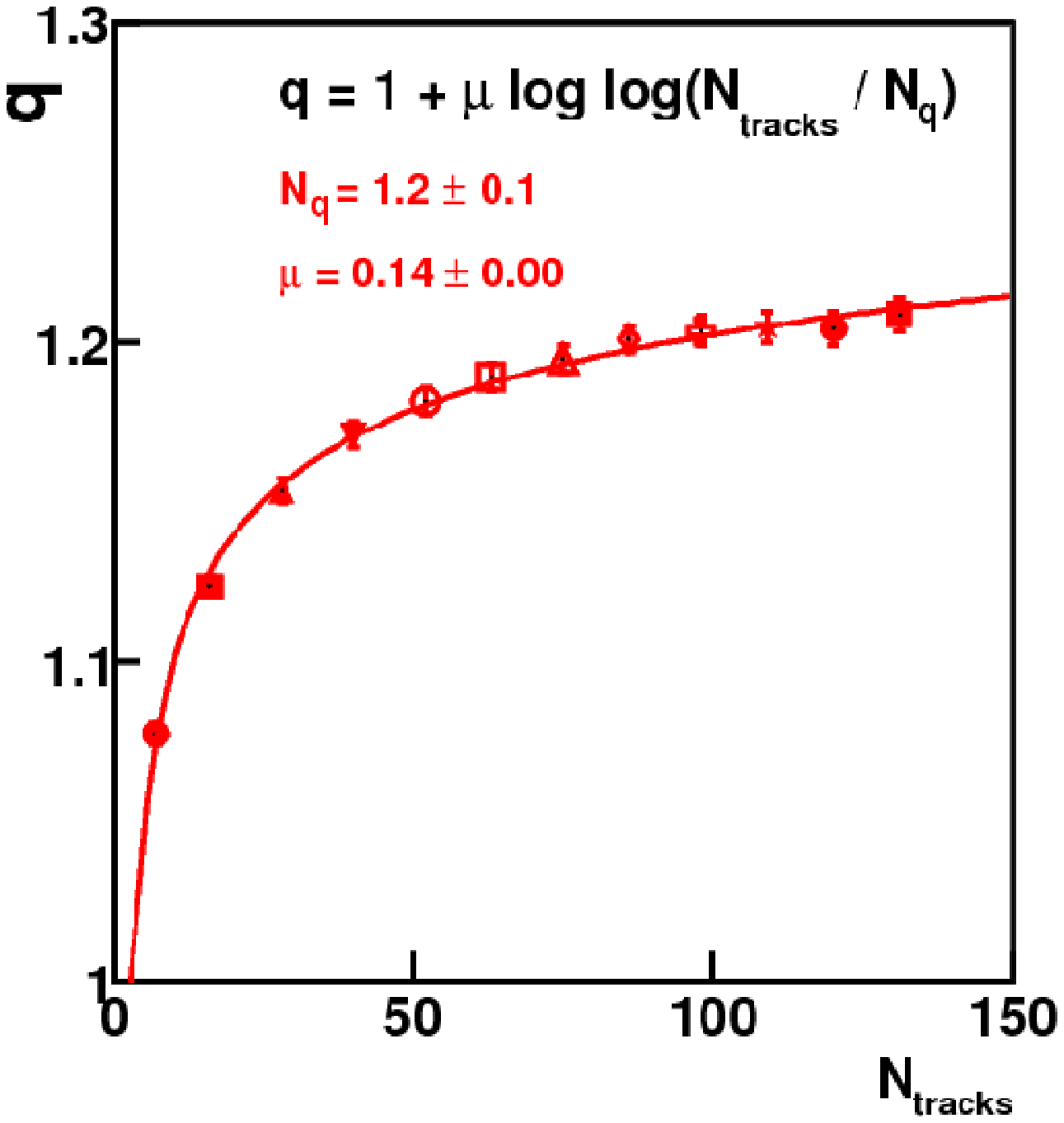}
\caption{Dependence of the $q$ parameter of Eq.~(\ref{spec1}) on the event-multiplicity obtained from fits to $\pi^+$ spectra shown in Fig.~\ref{fig:pipdNdpT}. Top: $\sqrt s$ = 0.9 TeV, middle: $\sqrt s$ = 2.76 TeV bottom: $\sqrt s$ = 7 TeV. Curves are fits of Eq.~(\ref{spec2}).  \label{fig:pipq}}
\end{figure}

\begin{figure}
\includegraphics[width=0.47\textwidth,height=0.32\textheight]{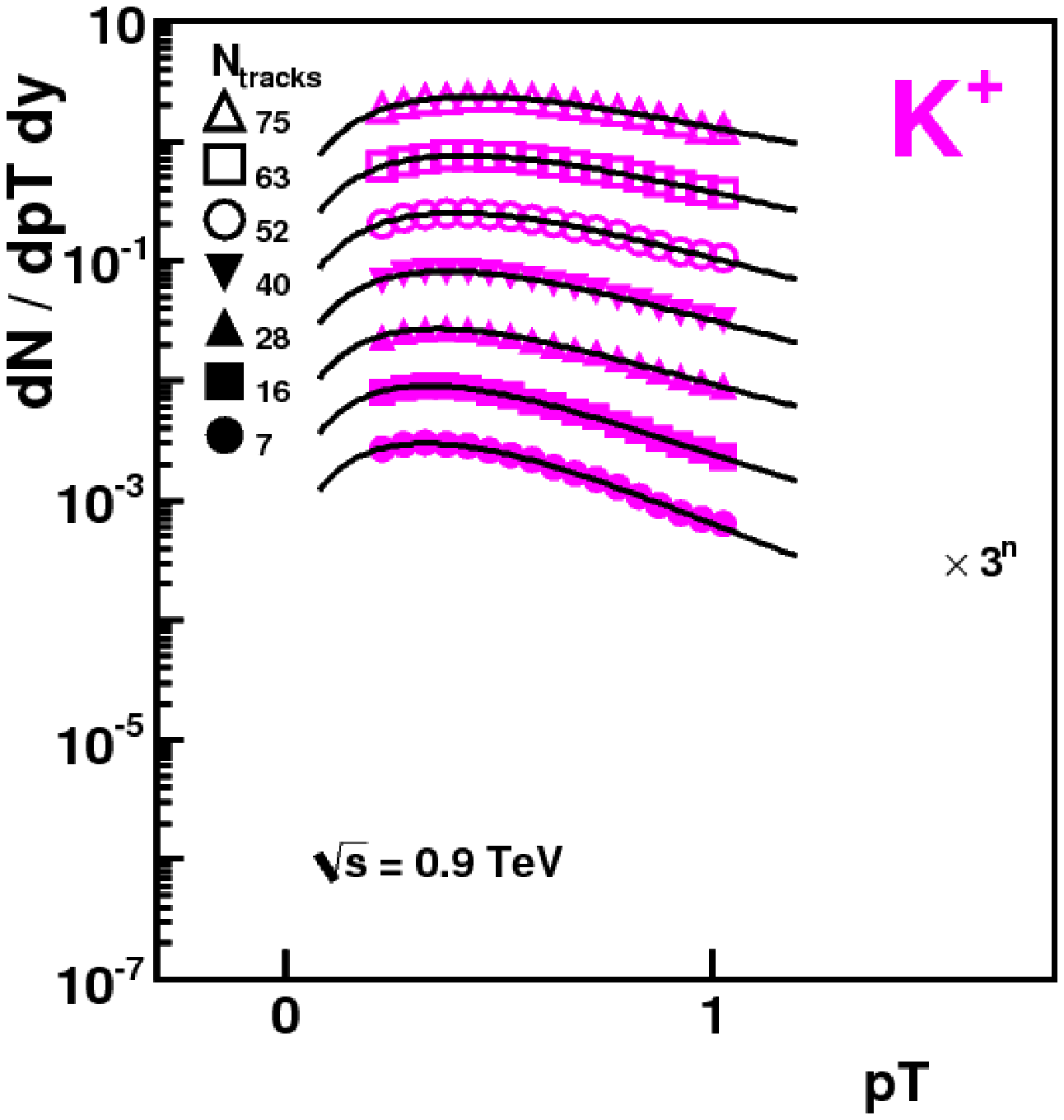}
\includegraphics[width=0.47\textwidth,height=0.32\textheight]{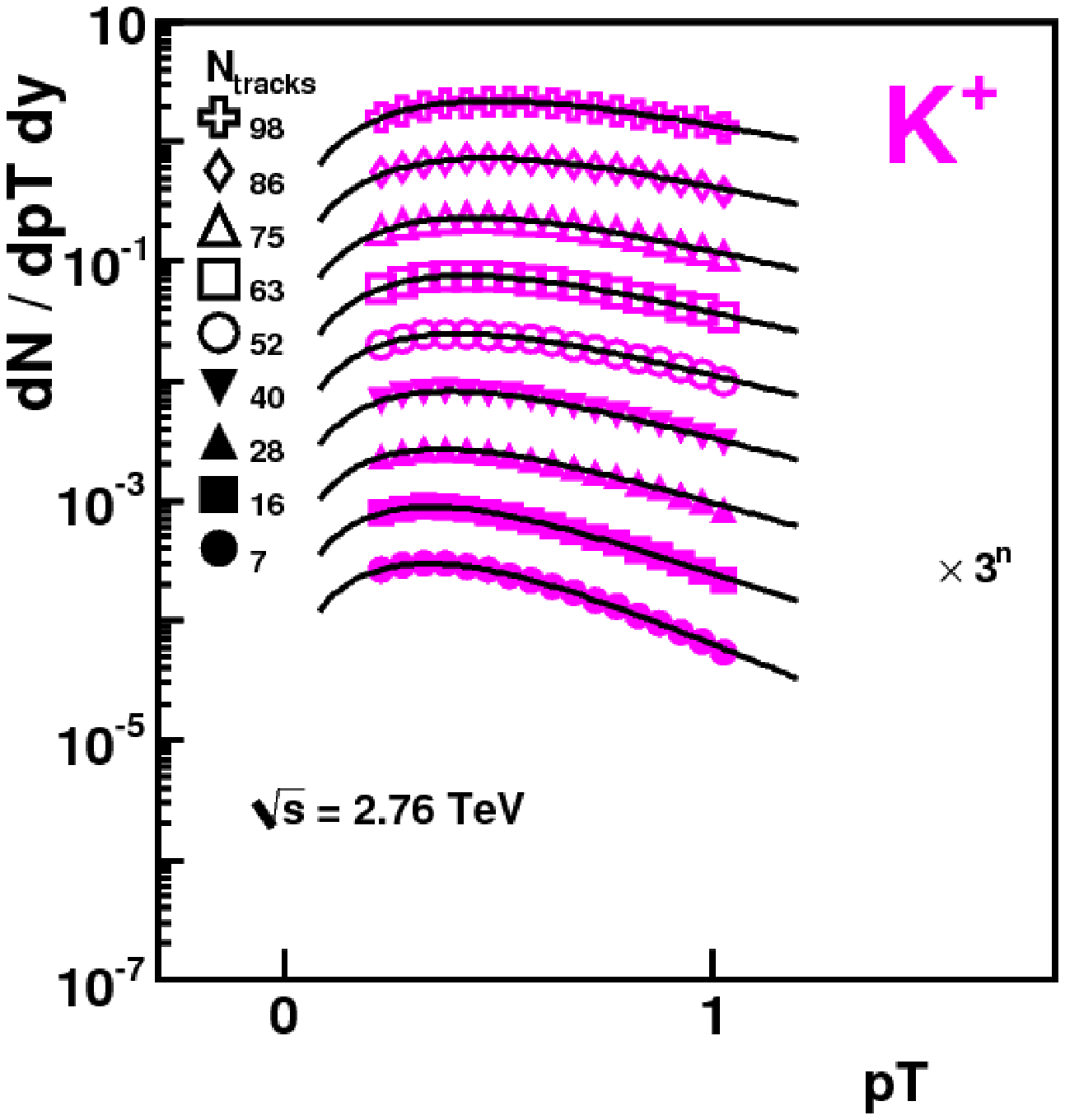}
\includegraphics[width=0.47\textwidth,height=0.32\textheight]{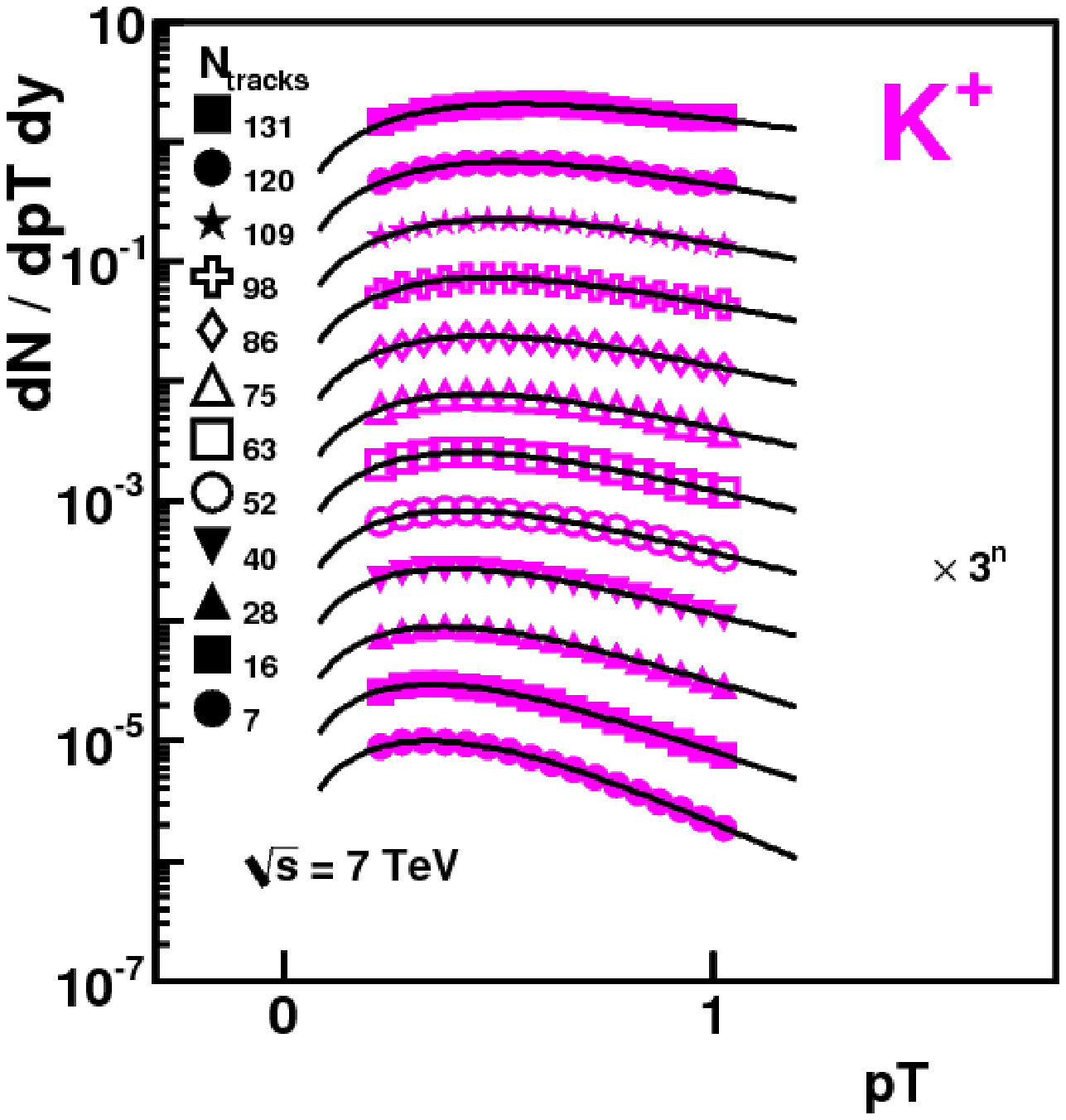}
\caption{Transverse momentum spectra of $K^+$-s stemming from $pp$ collisions of fix multiplicities, measured at $\sqrt{s}$ = 0.9 TeV (top), 2.76 TeV (middle) and 7 TeV (bottom) collision energies. Rapidity range: $|y|\leq1$. Data of graphs were published in \cite{bib:CMS}. Curves are fits of Eq.~(\ref{spec1}). \label{fig:KpdNdpT}}
\end{figure}

\begin{figure}
\includegraphics[width=0.47\textwidth,height=0.32\textheight]{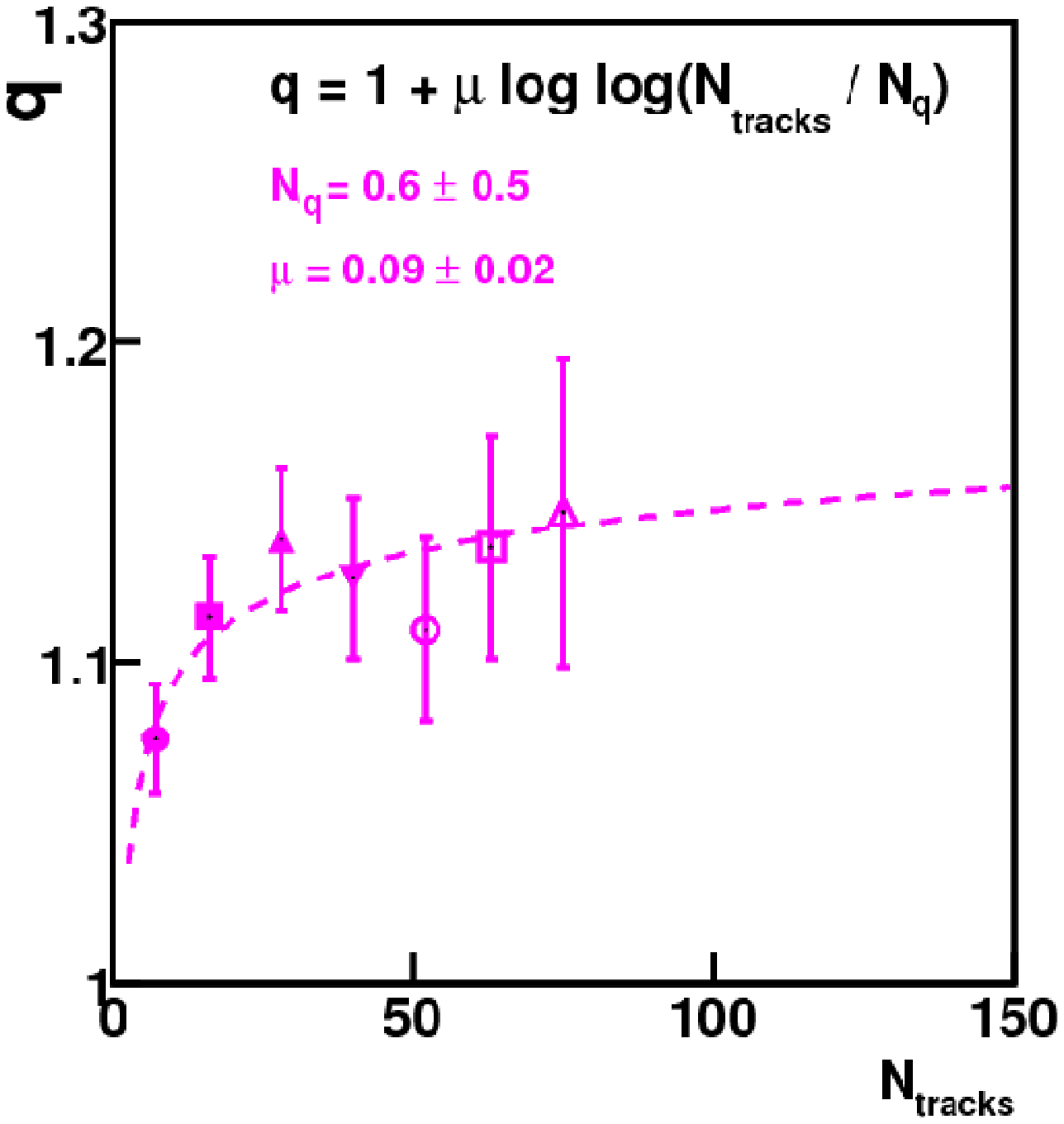}
\includegraphics[width=0.47\textwidth,height=0.32\textheight]{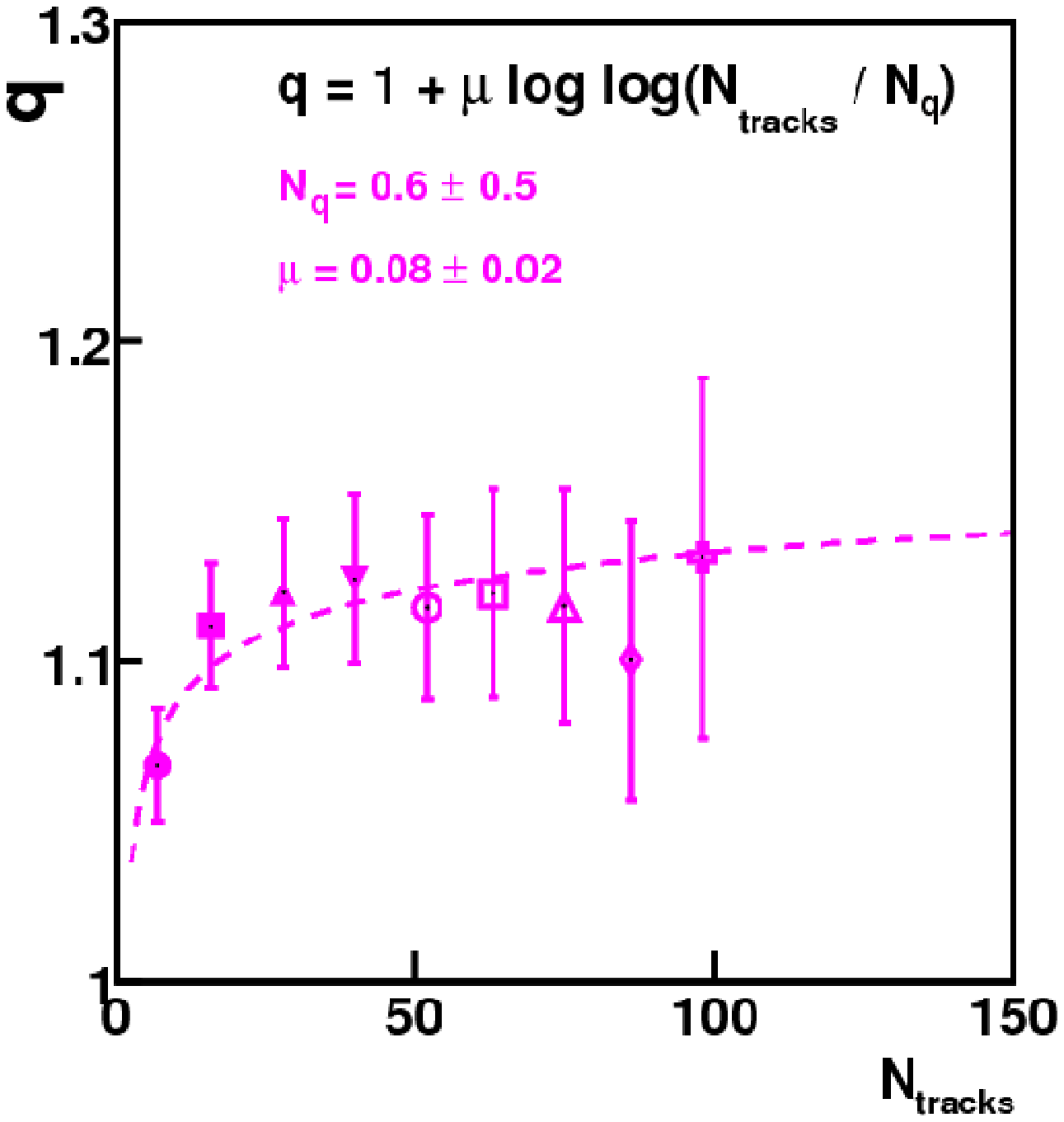}
\includegraphics[width=0.47\textwidth,height=0.32\textheight]{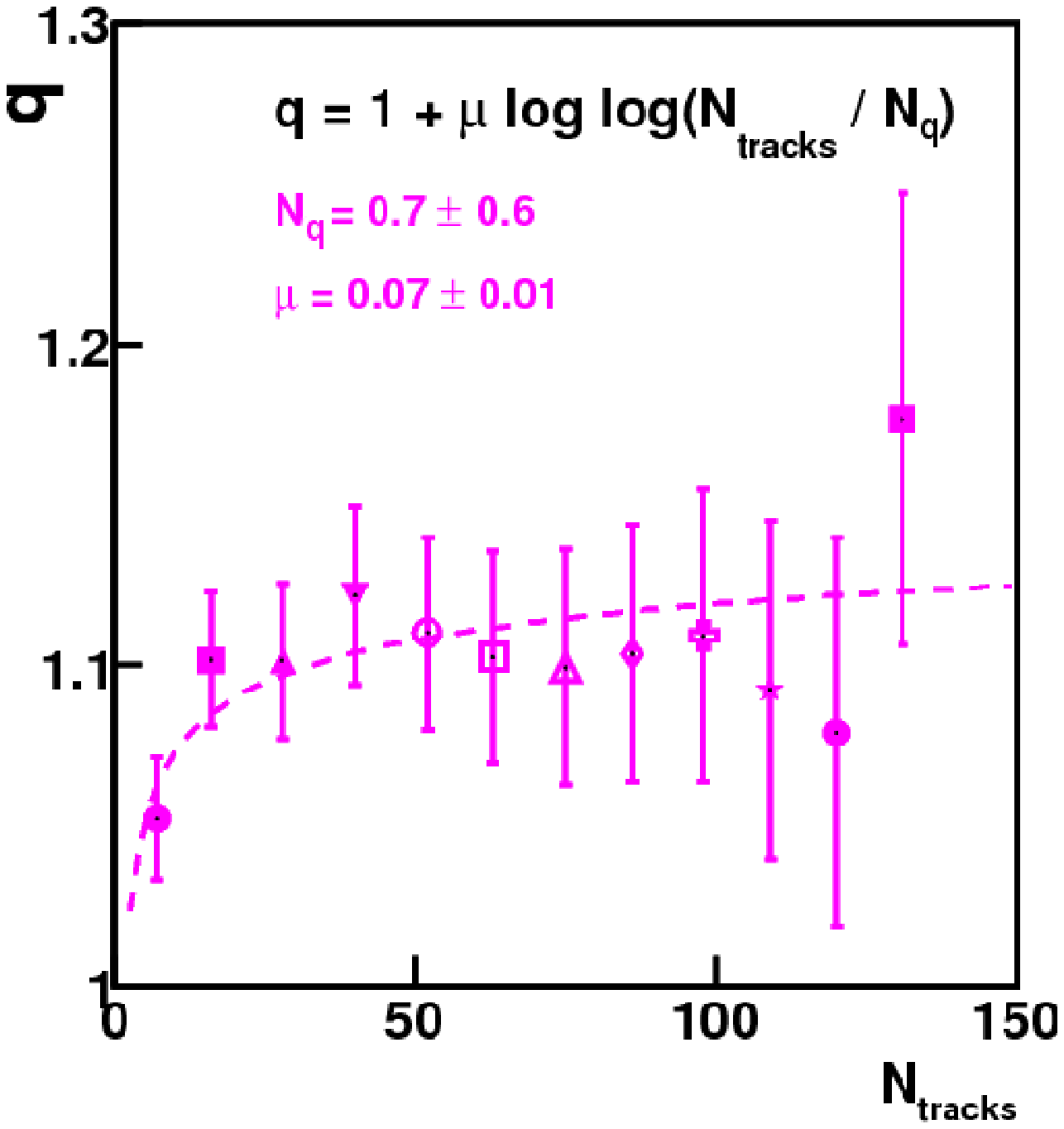}
\caption{Dependence of the $q$ parameter of Eq.~(\ref{spec1}) on the event-multiplicity obtained from fits to $K^+$ spectra shown in Fig.~\ref{fig:KpdNdpT}. Top: $\sqrt s$ = 0.9 TeV, middle: $\sqrt s$ = 2.76 TeV bottom: $\sqrt s$ = 7 TeV. Curves are fits of Eq.~(\ref{spec2}). \label{fig:Kpq}}
\end{figure}

\begin{figure}
\includegraphics[width=0.47\textwidth,height=0.32\textheight]{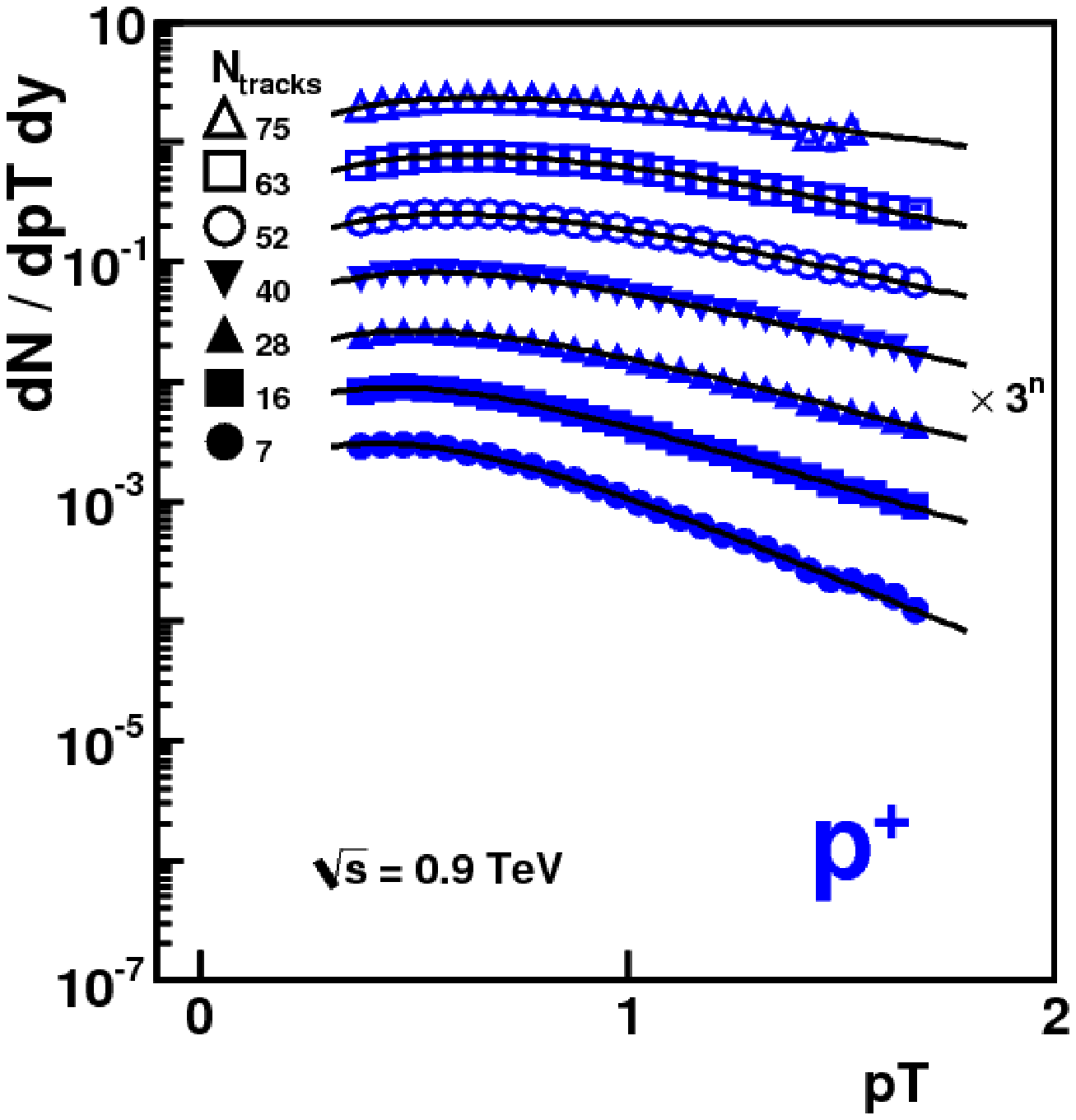}
\includegraphics[width=0.47\textwidth,height=0.32\textheight]{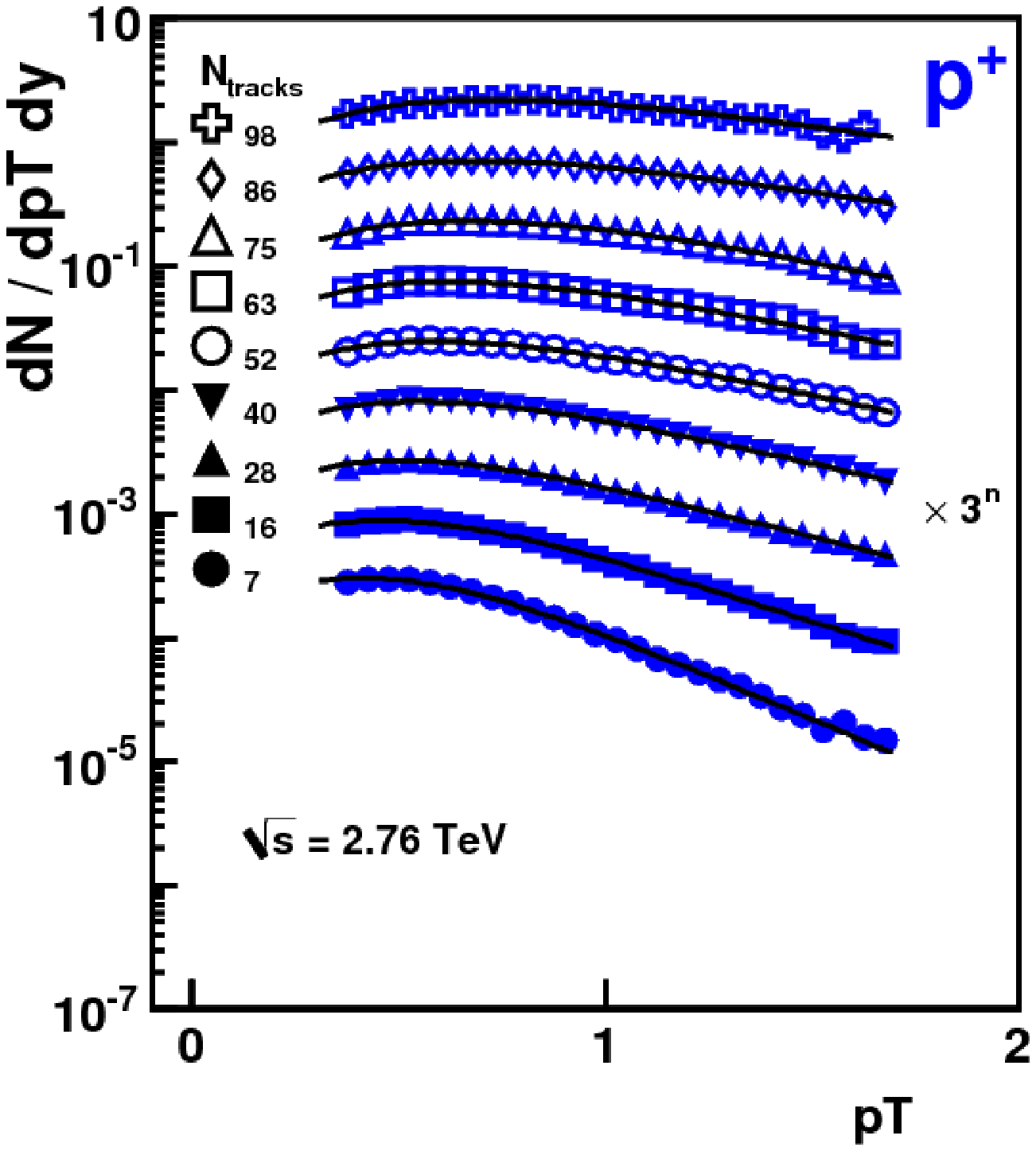}
\includegraphics[width=0.47\textwidth,height=0.32\textheight]{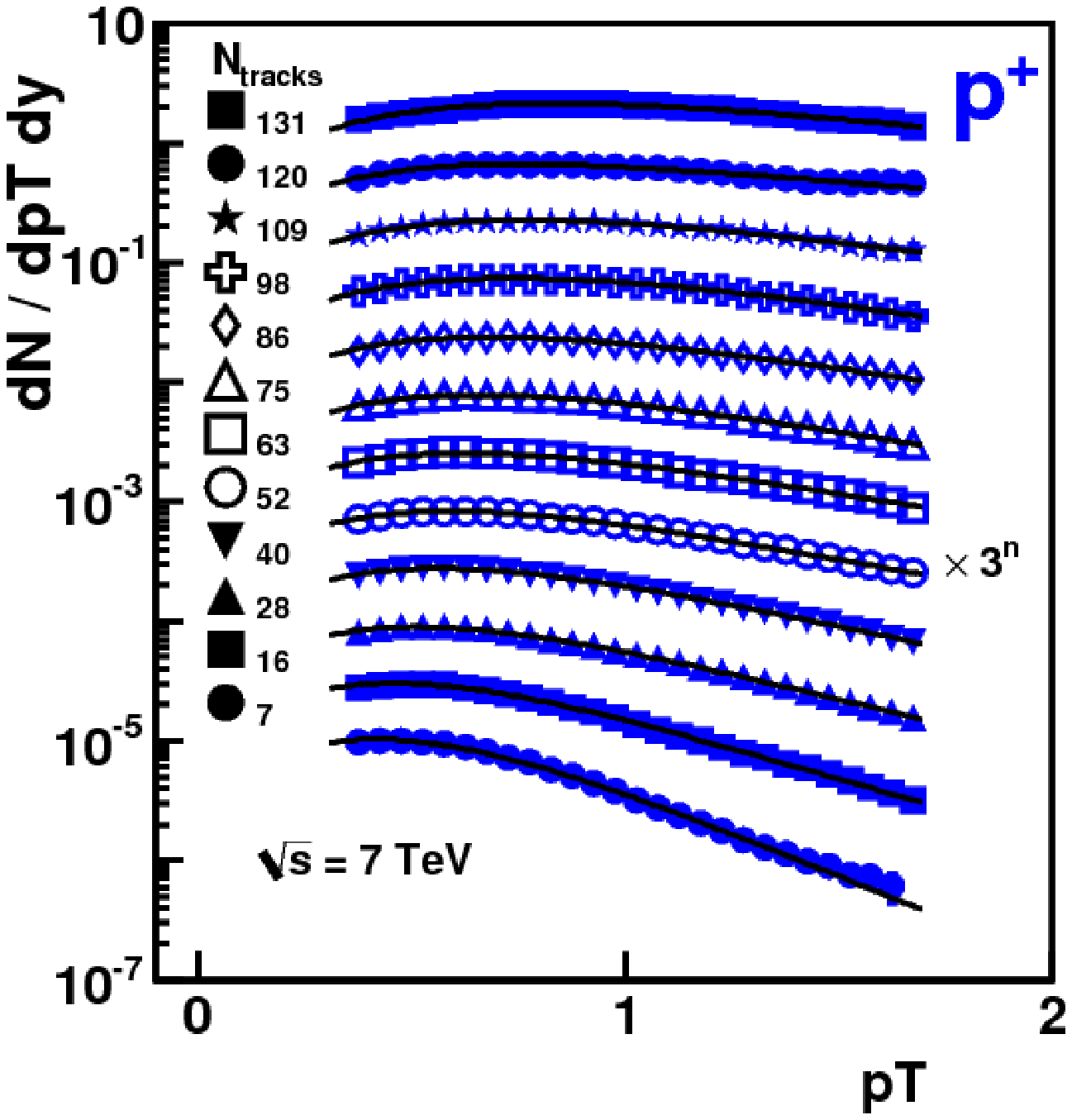}
\caption{Transverse momentum spectra of $p$-s stemming from $pp$ collisions of fix multiplicities, measured at $\sqrt{s}$ = 0.9 TeV (top), 2.76 TeV (middle) and 7 TeV (bottom) collision energies. Rapidity range: $|y|\leq1$. Data of graphs were published in \cite{bib:CMS}. Curves are fits of Eq.~(\ref{spec1}). \label{fig:ppdNdpT}}
\end{figure}

\begin{figure}
\includegraphics[width=0.47\textwidth,height=0.32\textheight]{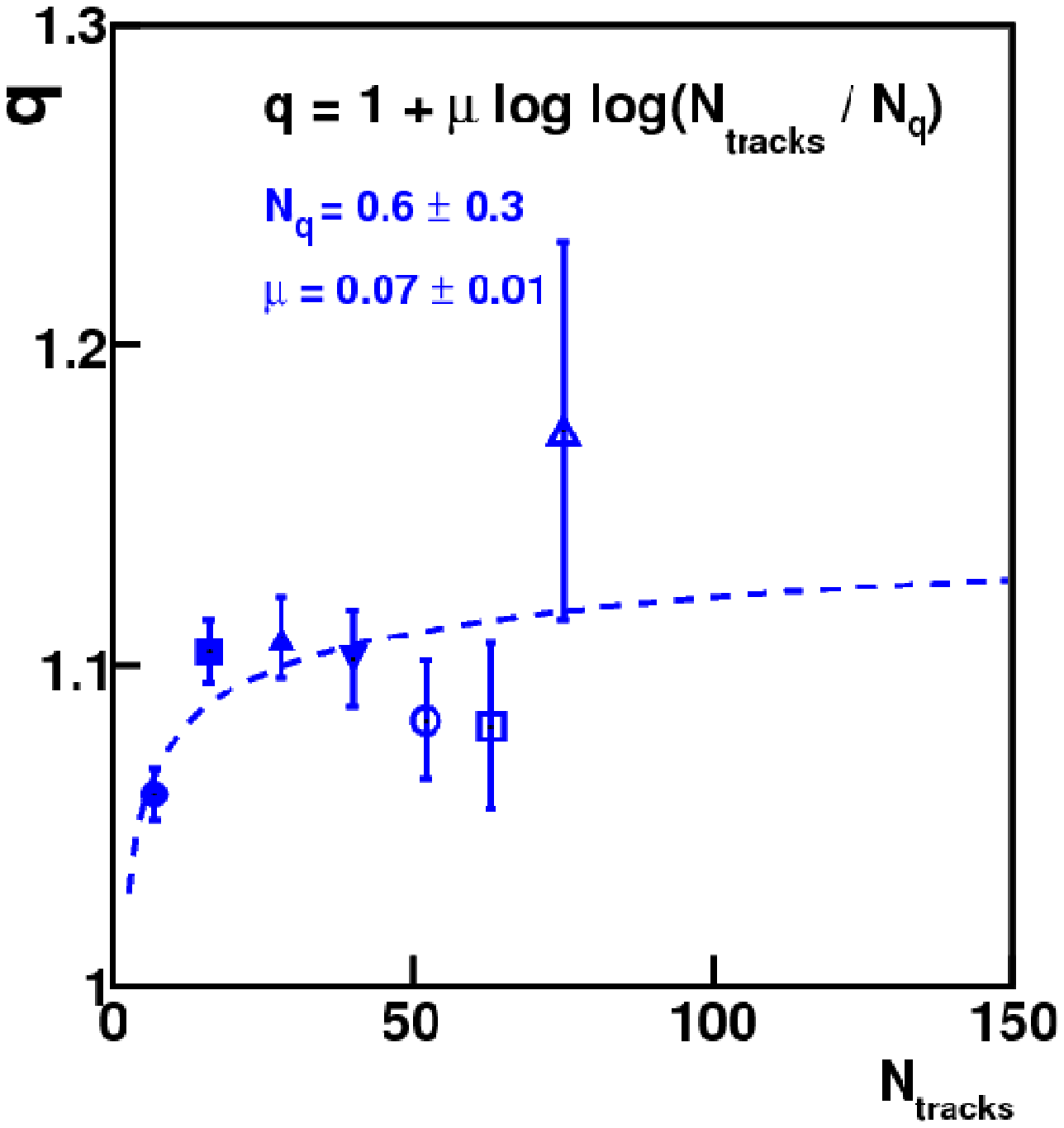}
\includegraphics[width=0.47\textwidth,height=0.32\textheight]{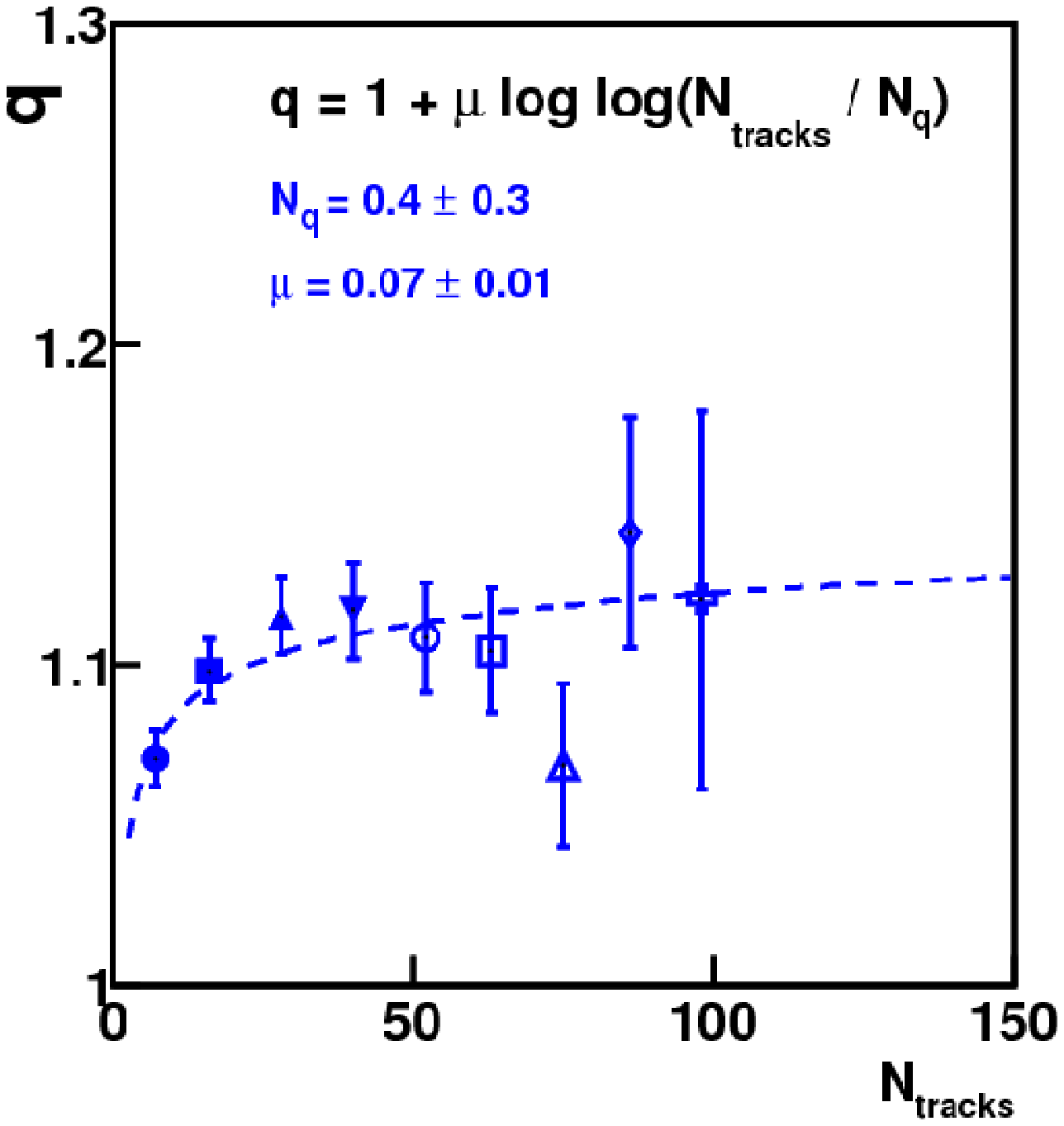}
\includegraphics[width=0.47\textwidth,height=0.32\textheight]{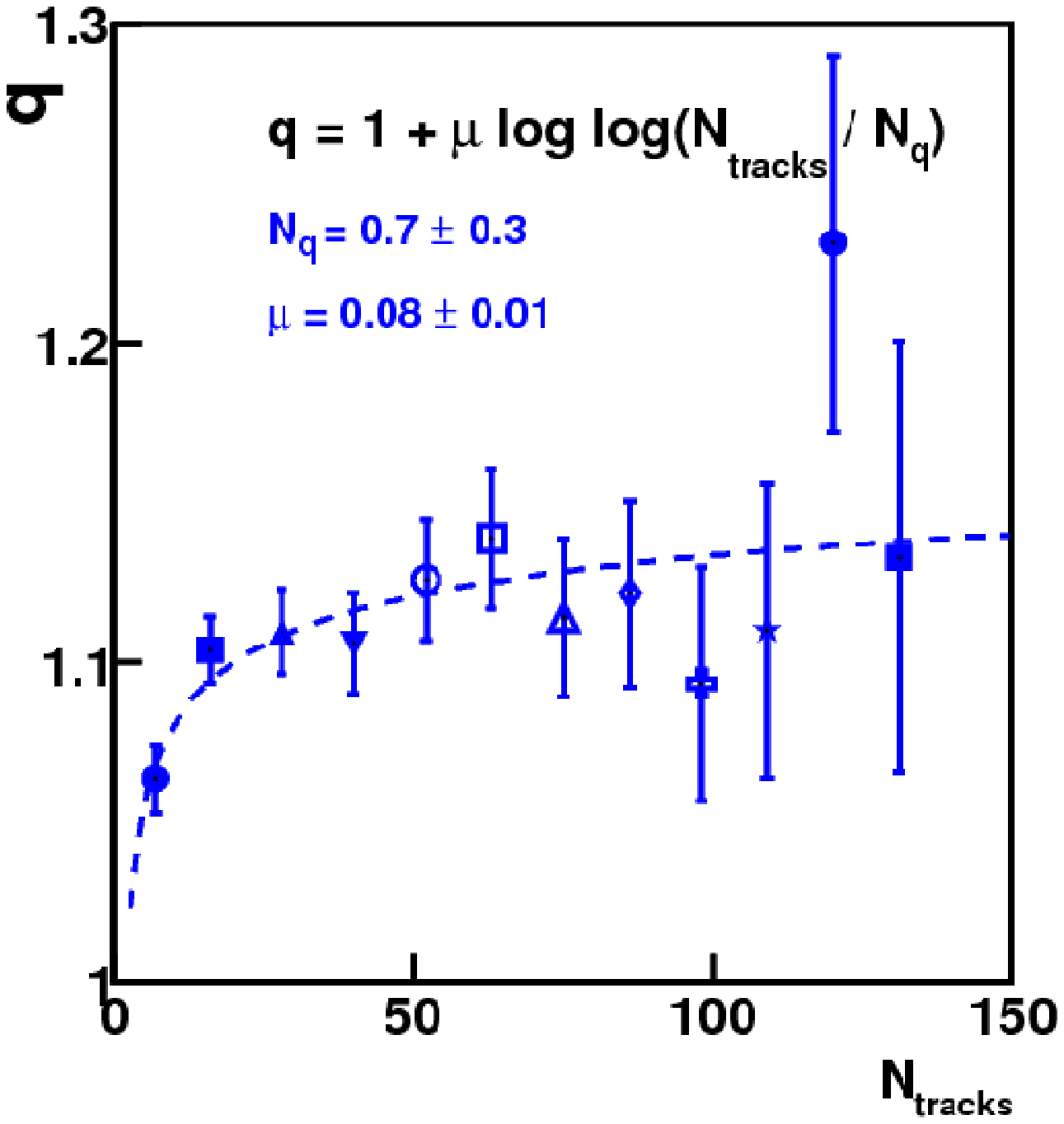}
\caption{Dependence of the $q$ parameter of Eq.~(\ref{spec1}) on the event-multiplicity obtained from fits to $p$ spectra shown in Fig.~\ref{fig:ppdNdpT}. Top: $\sqrt s$ = 0.9 TeV, middle: $\sqrt s$ = 2.76 TeV bottom: $\sqrt s$ = 7 TeV. Curves are fits of Eq.~(\ref{spec2}). \label{fig:ppq}}
\end{figure}

\begin{figure}
\includegraphics[width=0.47\textwidth,height=0.32\textheight]{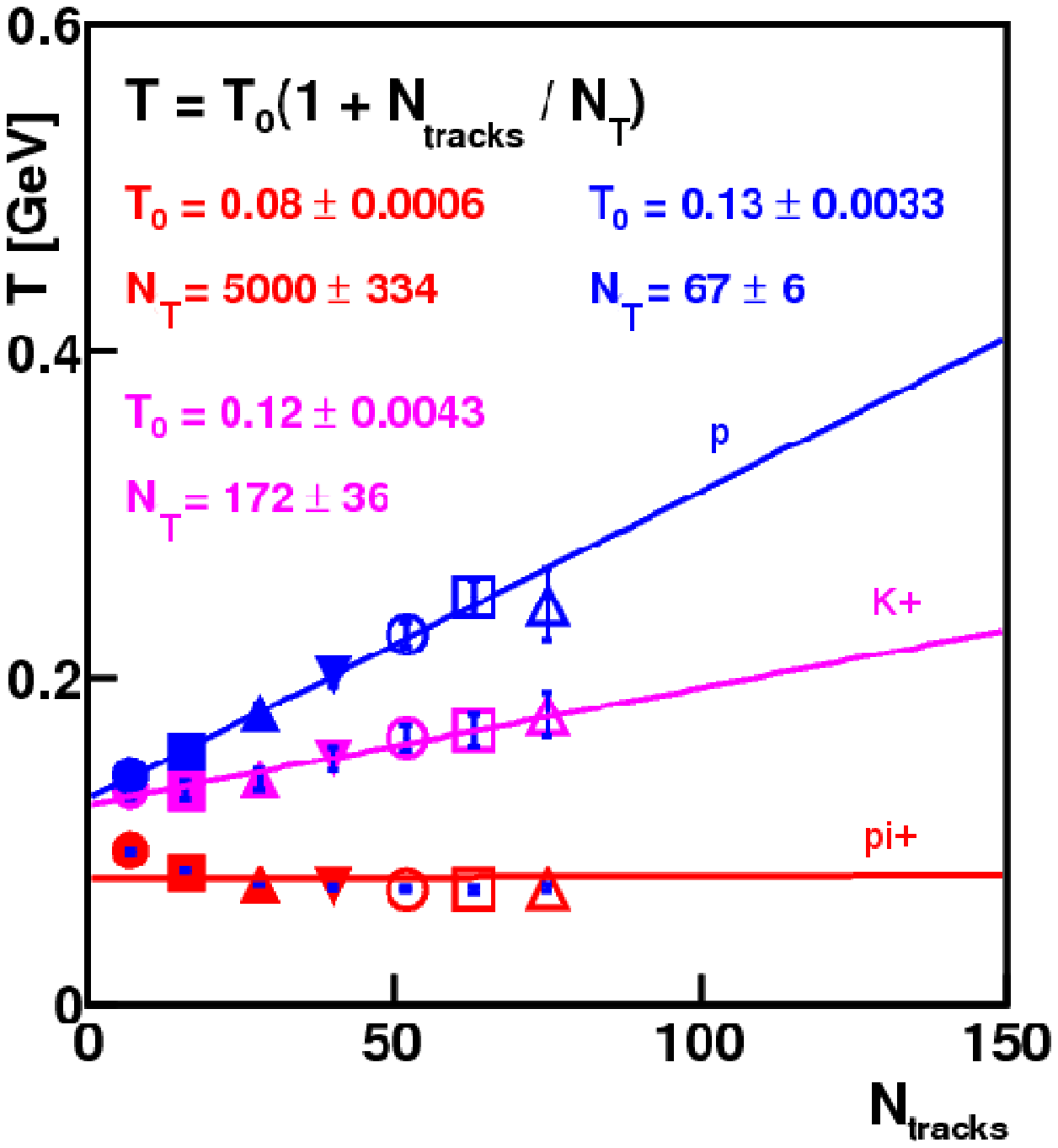}
\includegraphics[width=0.47\textwidth,height=0.32\textheight]{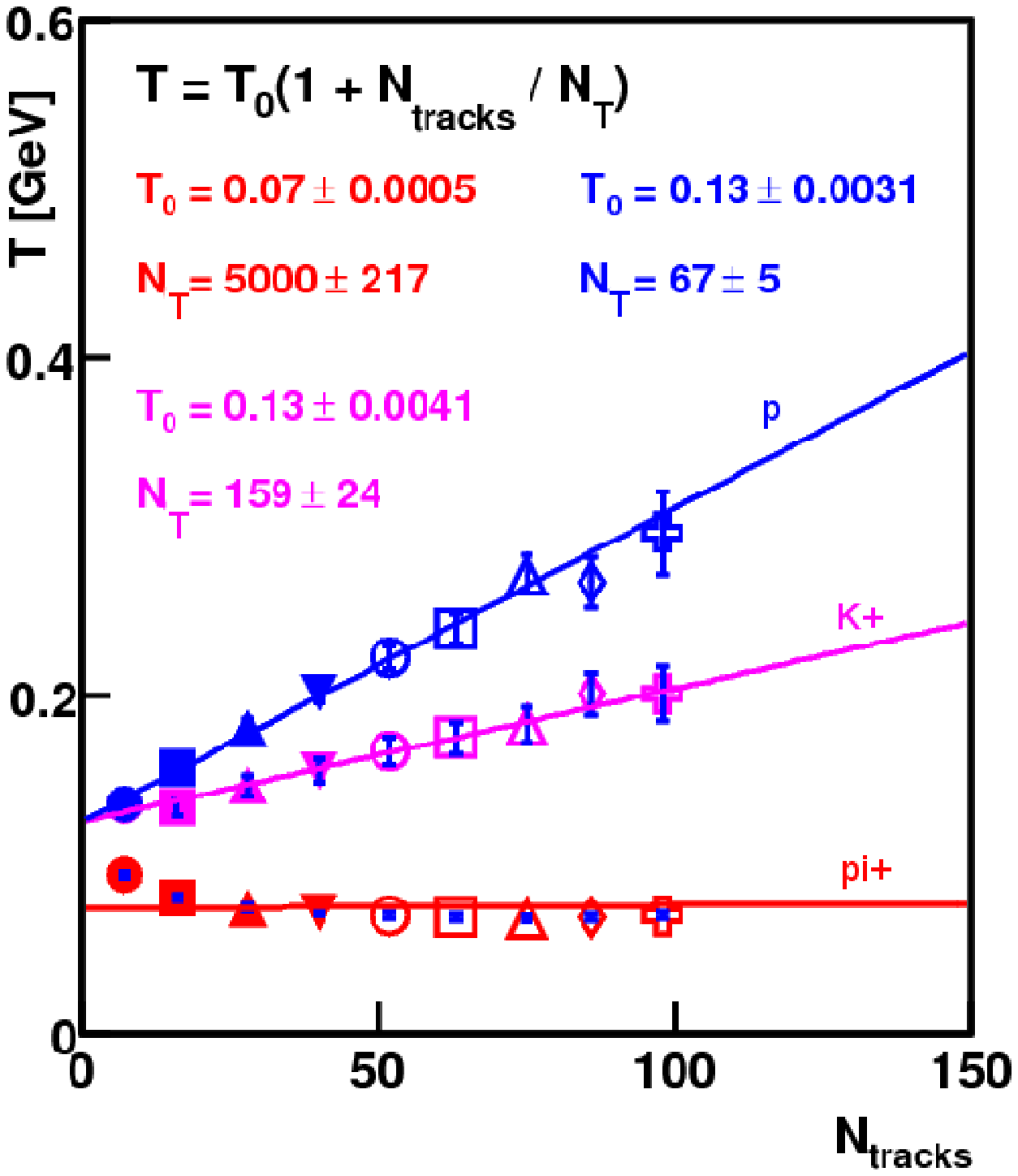}
\includegraphics[width=0.47\textwidth,height=0.32\textheight]{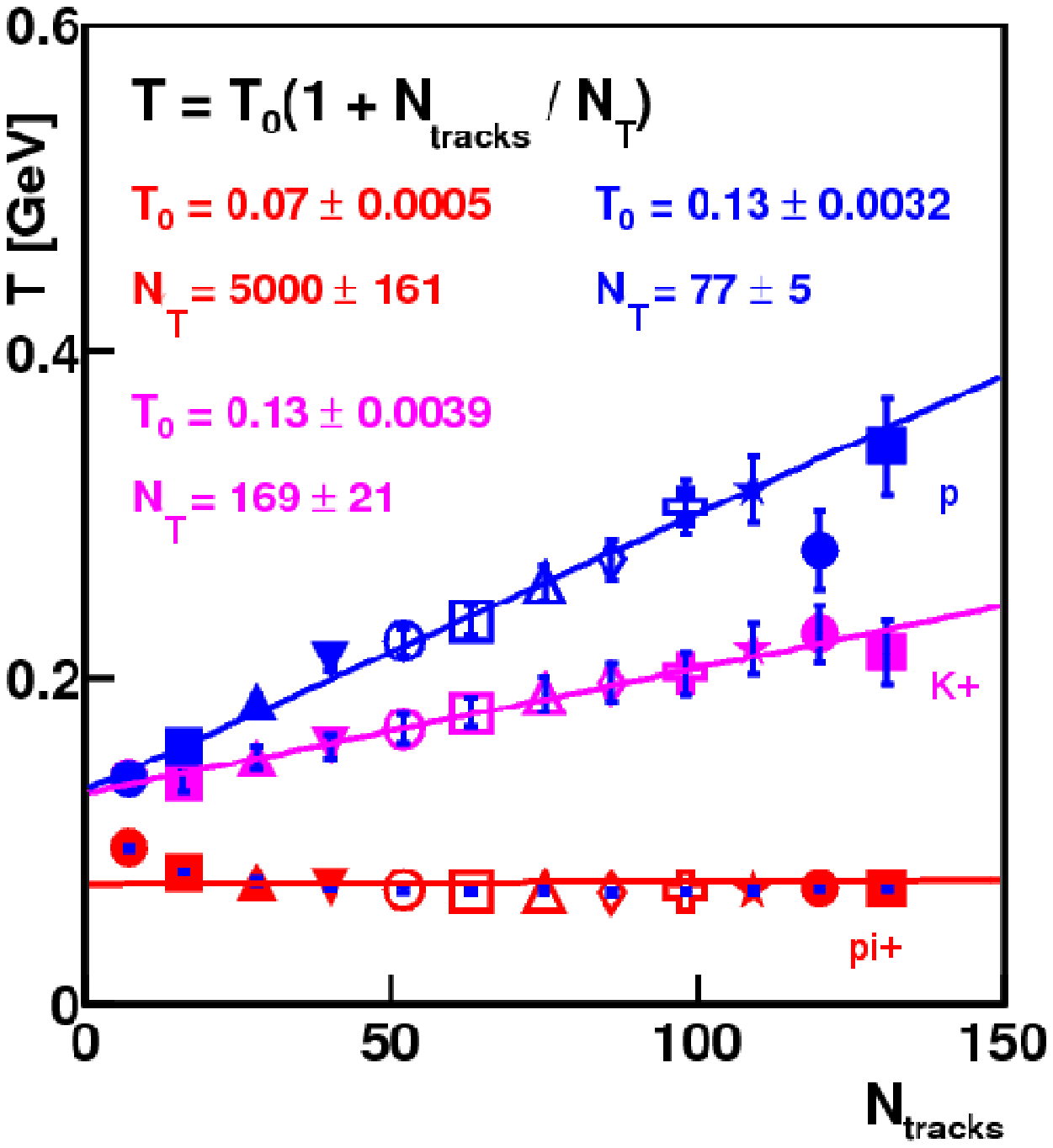}
\caption{Dependence of the $T$ parameter of Eq.~(\ref{spec1}) on the event-multiplicity obtained from fits to $\pi^+$, $K^+$ and $p$ spectra shown in Figs.~\ref{fig:pipdNdpT},~\ref{fig:KpdNdpT},~ and \ref{fig:ppdNdpT}. Top: $\sqrt s$ = 0.9 TeV, middle: $\sqrt s$ = 2.76 TeV bottom: $\sqrt s$ = 7 TeV. Curves are fits of Eq.~(\ref{spec2}). \label{fig:T}}
\end{figure}

The measurement of transverse spectra of a few types of hadrons stemming from proton--proton collisions for some fixed event multiplicities \cite{bib:CMS} made it possible to get rid of the effect of event-by-event multiplicity fluctuations on hadron spectra. In Sec.~\ref{sec:spec}, it is shown that transverse spectra of $\pi^+$, $K^+$ and $p$ take a cut-power law shape (Eq.~(\ref{spec1})) even in fixed multiplicity proton--proton events. It is also found that the $q$ parameter of the spectra follows a double-logarithmic dependence on the event-multiplicity $N$, while the $T$ parameter is independent of $N$ (within errors) for pions, but grows linearly with $N$ for kaons and protons (see Eq.~(\ref{spec2}) and Figs.~\ref{fig:pipdNdpT}~-~\ref{fig:T}). 

\begin{figure}
\includegraphics[width=0.47\textwidth,height=0.32\textheight]{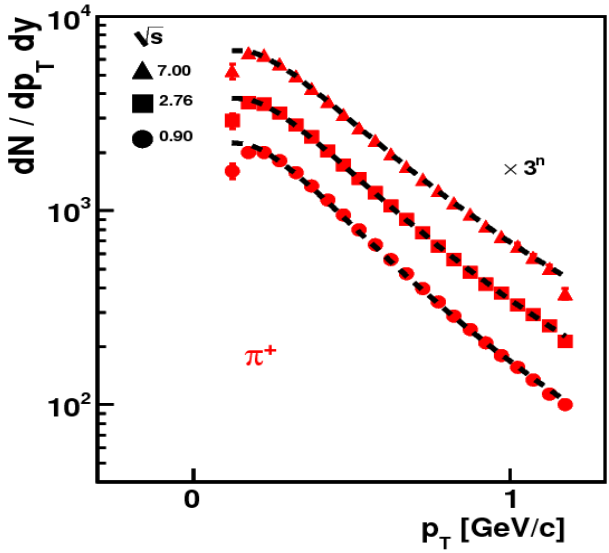}
\caption{Transverse momentum spectra of $\pi^+$-s stemming from $pp$ collisions and averaged over multiplicity fluctuations. Measured data are published in \cite{bib:CMS}. Dashed lines show Eq.~(\ref{sup8}) with different parametrisations of the multiplicity distribution Eq.~(\ref{sup5}). For collision energies $\sqrt s$ = 0.9, 2.76, and 7 TeV, $a = $ 1.38, 1.4, 1.12 and $N_0=$ 15.3, 24, 32 were used respectively (see text below Eq.~(\ref{sup8})).\label{fig:Naverpip}}
\end{figure}

\begin{figure}
\includegraphics[width=0.47\textwidth,height=0.32\textheight]{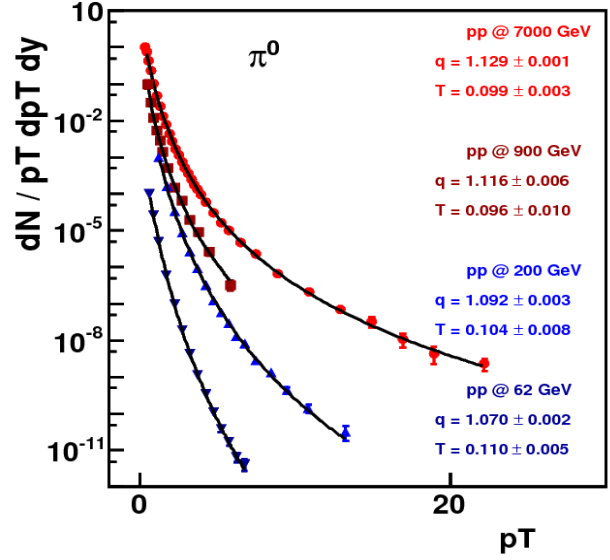}
\includegraphics[width=0.44\textwidth,height=0.30\textheight]{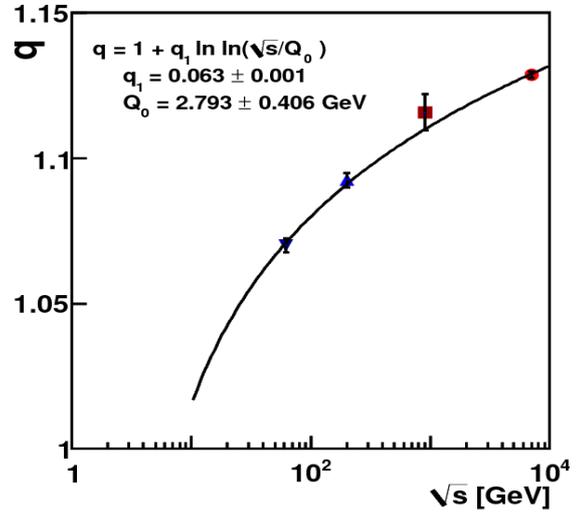}
\includegraphics[width=0.44\textwidth,height=0.30\textheight]{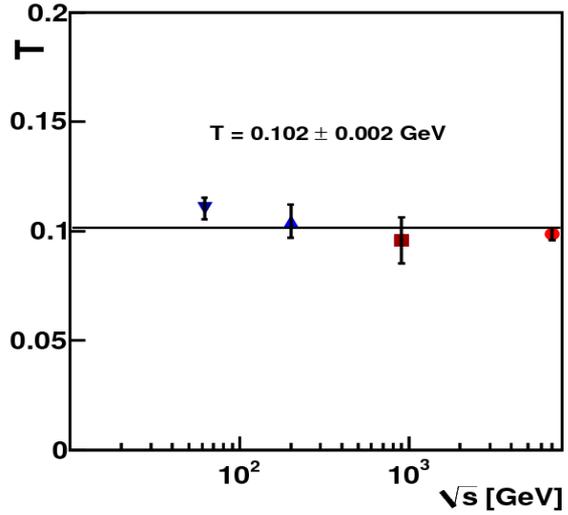}
\caption{Transverse momentum spectra of $\pi^0$-s stemming from $pp$ collisions at various collision energies and fits of Eq.~(\ref{spec1}) (top). $\sqrt{s}$ dependence of the $q$ parameter and fit of Eq.~(\ref{spec_1}) (middle). $\sqrt s$ dependence of the $T$ parameter (bottom) . Data of graphs were published in \cite{bib:ALICE3,bib:PHENIX1,bib:PHENIX2}.\label{fig:pi0}}
\end{figure}

In Sec.~\ref{sec:Efluct}, it is shown that cluster energy fluctuations of the form of Eq.~(\ref{sup4}) result in cut-power law shaped hadron spectrum at fix cluster multiplicity, if hadrons in a cluster are distributed according to the Boltzmann-Gibbs distribution. Averaging this spectrum over multiplicity fluctuations Eq.~(\ref{sup5}), using the weak multiplicity dependence of the $q$ and $T$ parameters of pion spectra found in Sec.~\ref{sec:spec}, also results in an approximately cut-power law spectrum that describes transverse $\pi^+$ spectrum (see Sec.~\ref{sec:ENfluct} and Fig.~\ref{fig:Naverpip}). This is consistent with the observation that multiplicity averaged transverse spectra (usually simply called 'transverse spectra') of hadrons stemming from $pp$ collisions fit the Tsallis distribution. In addition, I have also shown that the $q$ parameter of the transverse spectrum of $\pi^0$ shows a double-logarithmic dependence on the collision energy $\sqrt s$ in the range $\sqrt s \in$ [0.2, 7] TeV (see Fig.~\ref{fig:pi0}). Similar functional forms for $q(s)$ have also been proposed in \cite{bib:Wibig2,bib:Wibig,bib:Wilk3,bib:Wilk4}.

\begin{figure}
\includegraphics[width=0.47\textwidth,height=0.29\textheight]{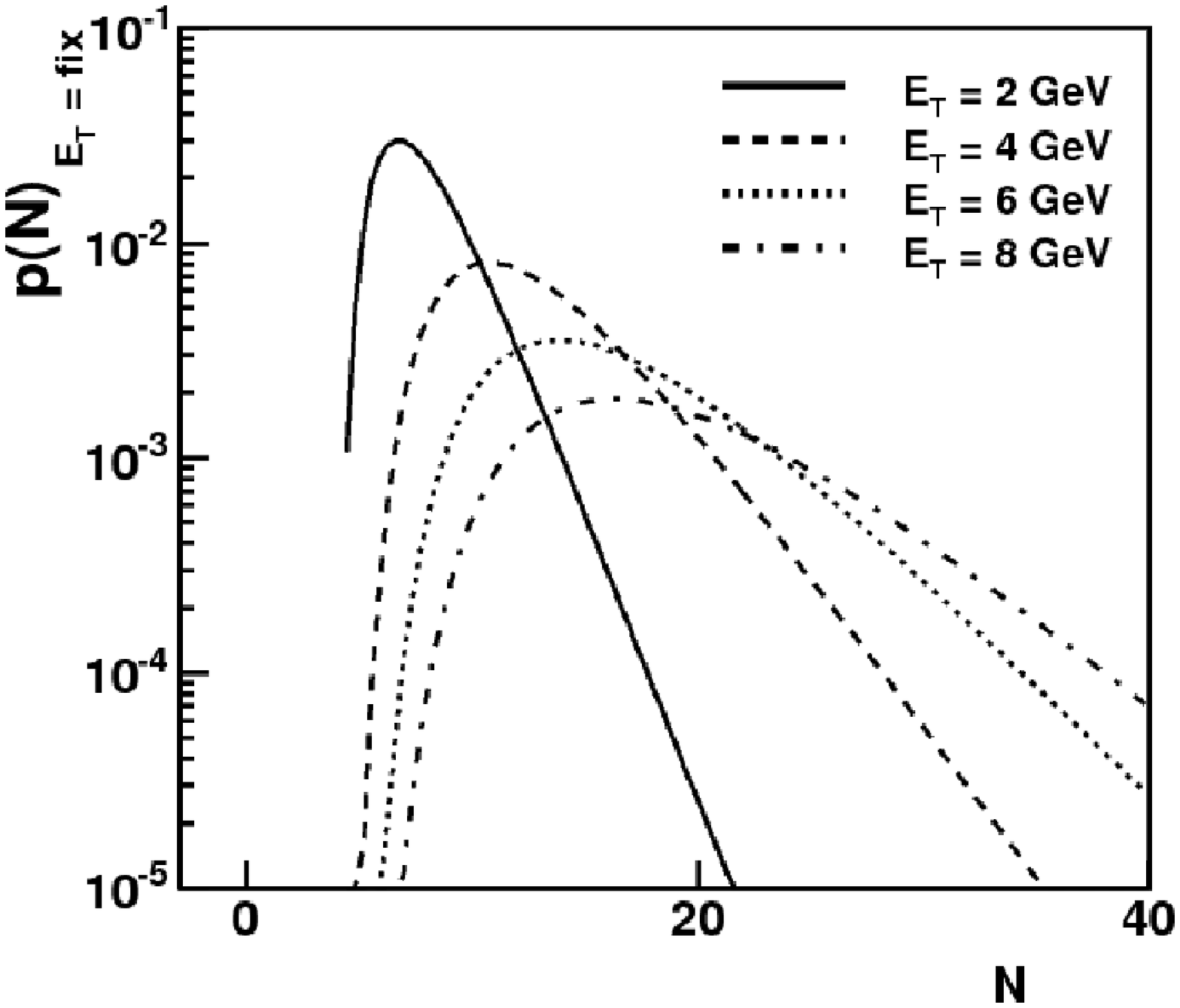}
\includegraphics[width=0.47\textwidth,height=0.29\textheight]{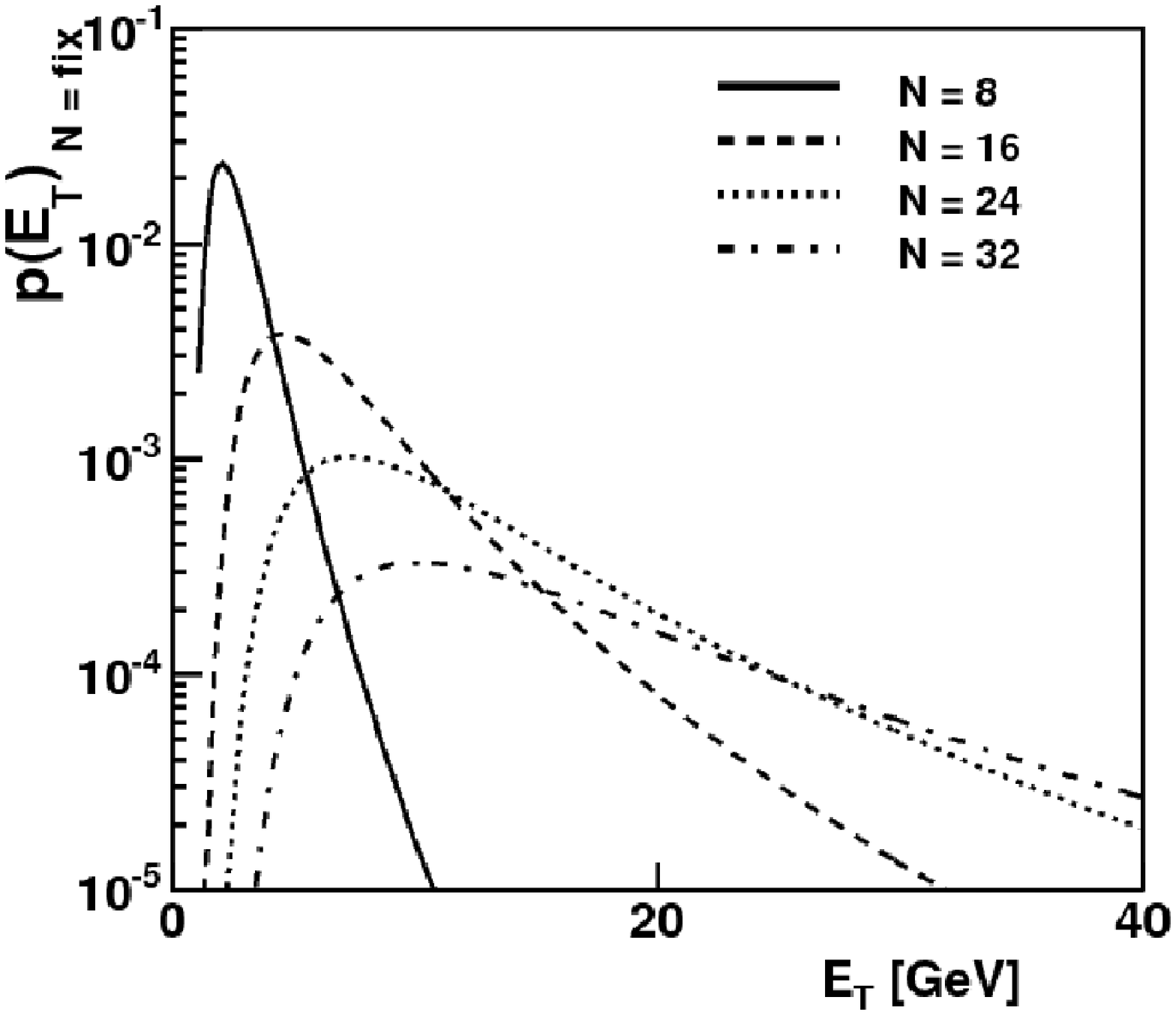}
\includegraphics[width=0.47\textwidth,height=0.29\textheight]{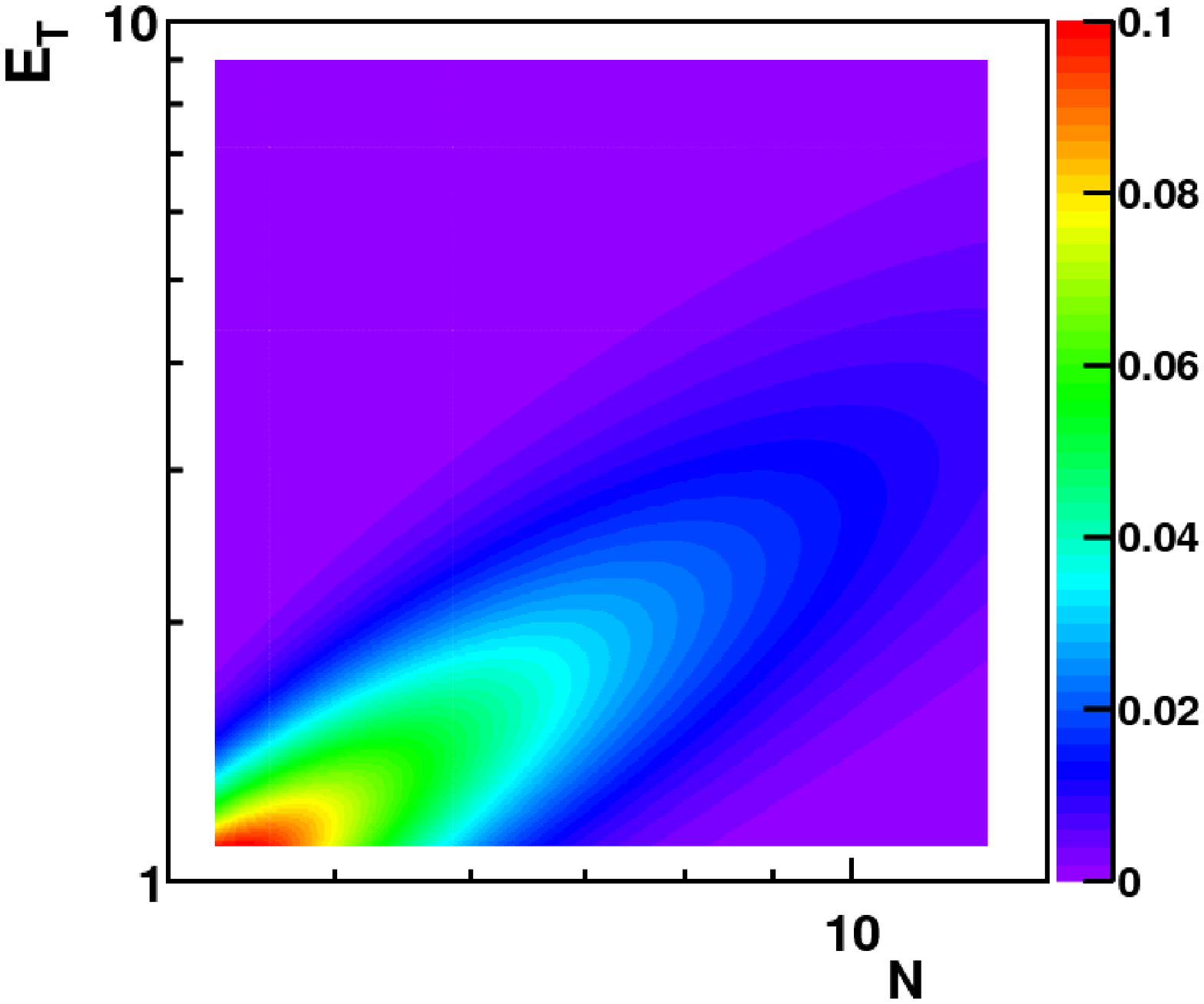}
\caption{Plots of Eq.~(\ref{pNE}) the joint distribution $p(N,E_T)$ of the event-multiplicity $N$ and total transverse energy $E_T$ (bottom) and its projections (top and middle). The parameters used were obtained from fits to the multiplicity distribution of charged hadrons and transverse $\pi^+$ spectra measured in $pp$ collisions at $\sqrt s$ = 7 TeV collision energy. Used parameters are: $a$ = 1.11, $N_0$ = 12, $\mu$ = 0.14, $N_q$ = 1.2, $T_0$ = 70 MeV, $N_T$ = 5000. \label{fig:pNE}}
\end{figure}

In order to be able to perform analytic calculations, I conjectured that only one cluster is formed in a single proton--proton event; hadrons stemming from the cluster have an isotropic Boltzmann-Gibbs distribution in the frame co-moving with the cluster; finally, I neglected particle masses and cluster velocities. In this ``first approximation'', the $E_c$ energy of the cluster for\-med in a single proton--proton event is proportional to the $E_T$ energy that reaches the transverse region ($|y|\leq1$ in \cite{bib:CMS}), $E_c\propto E_T$. Thus, the joint distribution $p(N,E_T)$ of the event multiplicity and the transverse energy in an event can be predicted (see Sec.~\ref{sec:pEN} and Fig.~\ref{fig:pNE}).

Though, it is to be emphasized that there are rough approximations in the above presented calculations (the case of multiple cluster production in a proton--proton event as well as non-zero cluster velocities and finite hadron masses are to be taken into account in future works), it is clearly pointed out that event-by-event fluctuation of the hadron multiplicity as well as that of the energy reaching the transverse range may result in cut-power law (or Tsallis) shaped hadron spectra. Even if hadrons had Boltzmann-Gibbs distribution in a single event or cluster.



%
%


\begin{acknowledgements}
This work was supported by the Hungarian OTKA Grant K104260. The author thanks Tamas S. Bir\'o, Gergely G. Barnaf\"oldi and Ferenc Sikl\'er for fruitful discussions.
\end{acknowledgements}



\end{document}